\documentclass[10pt]{llncs}

\usepackage[latin1]{inputenc}
\usepackage[T1]{fontenc}
\usepackage{float}
\usepackage{afterpage}
\usepackage{multicol}
\usepackage{amsmath}
\usepackage{amssymb}
\usepackage{amscd}
\usepackage{fancyhdr}
\usepackage{shadow}
\usepackage[usenames,dvipsnames]{xcolor}
\usepackage[ruled,vlined]{algorithm2e}
\usepackage{textcomp}
\usepackage{tabularx}

\usepackage{tikz}
\usetikzlibrary{snakes,backgrounds,arrows,matrix} 
\usepackage{extarrows}
\usepackage{fourier-orns}

\newtheorem{hyp}{Hypothesis}

\def\eex{\hfill\textreferencemark}
\def\ie{{\it i.e.}}

\def\eg{{\it e.g.}}

\def\t{{\ensuremath \Theta}}
\def\D{{\ensuremath {\cal D}}}
\def\Telem{{\cal T}_{\delta}^{\mbox{{\tiny elem}}}}
\def\GIG{{\sc gtg}}
\def\GGIG{\eff{\mbox{\GIG}}}
\def\SIG{{\sc atg}}
\def\SSIG{\eff{\mbox{\SIG}}}
\def\F{{\ensuremath {\cal F}}}
\def\G{{\ensuremath {\cal G}}}
\def\States{{\ensuremath X}}

\def\proba{\ensuremath{\mathbb{P}}}
\def\U{\ensuremath{\mathcal{U}}}
\def\nU{\ensuremath{\overline{\U}}}
\def\01{\ensuremath{\{0,1\}}}
\def\B{\ensuremath{\mathbb{B}}}
\def\Bn{\ensuremath{\B^n}}
\def\N{\ensuremath{\mathbb{N}}}

\newcommand{\NN}[1][n]{\ensuremath{\N/#1\N}}

\def\iff{{\ensuremath \ \Leftrightarrow\ }}

\def\Rar{{\ensuremath \ \Rightarrow\ }}
\def\mod{\mbox{ mod }}
\def\BSn{\ensuremath{\mbox{{\sc bs}}_n} }
\def\BSnH{\ensuremath{\widetilde{\mbox{{\sc bs}}}_n} }
\def\BSk[#1]{\ensuremath{\mbox{{\sc bs}}_{#1}} }
\def\BSnH{\ensuremath{\widetilde{\mbox{{\sc bs}}}_n} }
\def\BSkH[#1]{\ensuremath{\widetilde{\mbox{{\sc bs}}}_{#1}} }

\newcommand{\GTF}[1]{global transition function#1}
\newcommand{\us}[1]{update schedule#1}
\newcommand{\BS}[1]{block-sequential#1}
\newcommand{\BSus}[1]{block-sequential update schedule#1}
\newcommand{\PERus}[1]{periodic update schedule#1}

\newcommand{\Sus}[1]{simple update schedule#1}
\newcommand{\TG}[1]{transition graph#1}
\newcommand{\STS}[1]{state transition system#1}
\newcommand{\DS}[1]{dynamical system#1}
\newcommand{\BAN}[1]{Boolean automata network#1}
\newcommand{\LTF}[1]{local transition function#1}
\newcommand{\iieme}[1]{$#1^{\text{th}}$}

\newcommand{\PL}[0]{probability law}
\newcommand{\T}[1]{\ensuremath {{\cal T}_{#1}}}
\newcommand{\eff}[1]{\ensuremath{{#1}^{\mbox{{\tiny eff}}}}}
\newcommand{\trans}[1][]{
  \hspace{0.3pt}\raisebox{.5ex}{
    \begin{tikzpicture}[descr/.style={fill=white,inner sep=2.5pt}]
      \path (0,0) edge[->, >=angle 60] node[above]
      {\ensuremath{{\scriptstyle #1}}} (0.8,0);
    \end{tikzpicture}}~}
\newcommand{\Trans}[1][]{
  \hspace{0.3pt}\raisebox{.5ex}{
    \begin{tikzpicture}
      \path  (0,0) edge[-] (0.4,0);
    \end{tikzpicture}}\,
    \raisebox{0.15ex}{\ensuremath{{\scriptstyle #1}}}\,
    \raisebox{0.5ex}{\begin{tikzpicture}
      \path  (0,0) edge[->, >=angle 60] (0.5,0);
    \end{tikzpicture}}~}
\newcommand{\seq}[1][]{
  \hspace{0.3pt}\raisebox{.5ex}{
    \begin{tikzpicture}[descr/.style={fill=white,inner sep=2.5pt}]
      \path (0,0) edge[-open triangle 60] node[above]
      {\ensuremath{{\scriptstyle #1}}} (0.8,0);
    \end{tikzpicture}}~}
\newcommand{\Seq}[1][]{
  \hspace{0.3pt}\raisebox{.5ex}{
    \begin{tikzpicture}
      \path  (0,0) edge[-] (0.3,0);
    \end{tikzpicture}}\,
    \raisebox{0.15ex}{\ensuremath{{\scriptstyle #1}}}\,
    \raisebox{0.5ex}{\begin{tikzpicture}
      \path  (0,0) edge[-open triangle 60] (0.5,0);
    \end{tikzpicture}}~}
\newcommand{\pll}[1][]{
  \hspace{0.3pt}\raisebox{.5ex}{
    \begin{tikzpicture}[descr/.style={fill=white,inner sep=2.5pt}]
      \path (0,0) edge[-triangle 60] node[above]
      {\ensuremath{{\scriptstyle #1}}} (0.8,0);
    \end{tikzpicture}}~}
\newcommand{\Pll}[1][]{
  \hspace{0.3pt}\raisebox{.5ex}{
    \begin{tikzpicture}
      \path  (0,0) edge[-] (0.4,0);
    \end{tikzpicture}}\,
    \raisebox{0.15ex}{\ensuremath{{\scriptstyle #1}}}\,
    \raisebox{0.5ex}{\begin{tikzpicture}
      \path  (0,0) edge[-triangle 60] (0.5,0);
    \end{tikzpicture}}~}
\newcommand{\transRT}[1][]{
  \hspace{0.3pt}\raisebox{.5ex}{
    \begin{tikzpicture}[descr/.style={fill=white,inner sep=2.5pt}]
      \path[font=\scriptsize] (0,0) edge[->>, >=angle 60] node[above]
      {\ensuremath{{\scriptstyle #1}}} (0.8,0);
    \end{tikzpicture}}~}
\newcommand{\REVtransRT}[1][]{
  \hspace{0.3pt}\raisebox{.5ex}{
    \begin{tikzpicture}[descr/.style={fill=white,inner sep=2.5pt}]
      \path (0,0) edge[<<-, >=angle 60] node[above]
      {\ensuremath{{\scriptstyle #1}}} (0.8,0);
    \end{tikzpicture}}~}

\def\doubleHDarr{\raisebox{.6ex}{
    \begin{tikzpicture}
      \draw[->>, >=angle 60](0,0)--(2.5,0);
    \end{tikzpicture}}}
\newcommand{\transRTW}[1][]{{\ensuremath \hspace{0.5pt}
    \stackrel{#1}{\doubleHDarr}\hspace{1pt}}}
\newcommand{\transOBS}{{\ensuremath \hspace{0.5pt} \raisebox{.3ex}{
      \begin{tikzpicture}
        \draw[-,snake=bumps,segment aspect=-2](0,0)--(0.7,0); 
        \draw[->, >=angle 60,](0.7,0)--(0.85,0);
    \end{tikzpicture}} \hspace{1.5pt}}}
\newcommand{\tip}[1][i]{\ensuremath{d_{#1}}}

\newcommand{\delPG}[2][1]{{\ensuremath d_{#2/#1}}} 
\def\del{\ensuremath{d}}
\newcommand{\delup}[1][i]{\ensuremath{d_{#1}^+}}
\newcommand{\deldwn}[1][i]{\ensuremath{d_{#1}^-}}
\newcommand{\PGPG}[4]{{\ensuremath
    \left[\begin{array}{@{\hspace{3pt}}c@{\hspace{4pt}}c@{\hspace{3pt}}} #1 &
    #3\\#2 & #4 \end{array}\right]} }
\newcommand{\PG}[2]{{\ensuremath
    \left[\begin{array}{@{\hspace{3pt}}c@{\hspace{3pt}}} #1
        \\#2\end{array}\right]}}

\newcommand{\fcn}[4]{\ensuremath{\left\{\begin{array}{rcl} #1 & \to & #2\\#3 &
        \mapsto & #4\end{array}\right.}}

\begin{document}

\pagestyle{plain}

\title{Towards a theory of modelling with Boolean automata networks -- I. Theorisation and observations}

\author{Mathilde Noual\inst{1,3} \and Sylvain Sené\inst{2,3}}

\institute{
  Universit{\'e} de Lyon, {\'E}NS-Lyon, LIP, CNRS UMR 5668, 69007 Lyon,
  France\and
  Universit{\'e} d'{\'E}vry -- Val d'Essonne, IBISC, {\'E}A 4526, 91000
  {\'E}vry, France\and
  Institut rh{\^o}ne-alpin des syst{\`e}mes complexes, IXXI, 69007 Lyon,
  France
}

\date{}

\maketitle

\begin{abstract}
Although models are built on the basis of some observations of reality, the
concepts that derive theoretically from their definitions as well as from their
characteristics and properties are not necessarily direct consequences of these
initial observations. Indeed, many of them rather follow from chains of
theoretical inferences that are only based on the precise model definitions and
rely strongly, in addition, on some consequential working hypotheses. Thus, it
is important to address the question of which features of a model effectively
carry some modelling meaning and which only result from the task of formalising
observations of reality into a mathematical language. In this article, we
address this question with a theoretical point view that sets our discussion
strictly between the two stages of the modelling process that require knowledge
of real systems, that is, between the initial stage that chooses a global
theoretical framework to build the model and the final stage that exploits its
formal predictions by comparing them to the reality that the model was designed
to simulate. Taking Boolean automata networks as instances of models of systems
observed in reality, we analyse in this setting the remaining stages of the
modelling process and we show how the meaning of theoretical concepts can subtly
rely on formal choices such as definitions and hypotheses.
\end{abstract}


\vspace{2cm}

\section{Introduction}
\label{SEC-introduction} 

The manipulation of mathematical objects requires the design and elaboration of
precise formal definitions in relation to these objects and to their properties.
Choosing these definitions, specifying consistent connections between them as
well as, possibly, restricting and refining them \emph{a posteriori} depends
closely on the original purpose of the objects that are considered. Some
theoretical objects such as \BAN{s} may be regarded and studied plainly as
mathematical objects that are disconnected from any modelling
considerations~\cite{vonNeumann1966,Gardner1970,Goles1990,Robert1995}. And they
can also be considered as models of other systems, possibly more complex systems
such as those encountered and observed in
reality~\cite{McCulloch1943,Kauffman1969,Schelling1971,Thomas1973}. Of course,
the approaches that follow from these two viewpoints are not independent. On the
one hand, formal studies may -- and sometimes need to -- be fed by practical
interrogations that arise from more applicative
contexts~\cite{Ising1925,Mandelbrot1983}. On the other hand, by definition,
modelling aims at inferring properties of a real system by exploiting a knowledge
of the properties of its mathematical model~\cite{Thom1989,Mendoza1999}.  Yet,
with the first point of view, the mathematical objects are studied \emph{per
  se}. The choices of formalisation and, more generally, the methodology that is
adopted mainly aim at building or expanding a purely theoretical understanding
in the domain at hand. In particular, possible restrictions to the scope of a
study are usually brought for the simple sake of convenience in formal
developments. On the contrary, the second point of view introduces motivations
of a different nature. Indeed, with the ultimate aim of simulating or explaining
a portion of reality, the choices of formalisation become oriented both by a
predefined or intuitive interpretation of mathematical notions and by,
conversely, the intention of representing given features of reality. As a
consequence, the second point of view yields a notable difficulty of which the
first purely formal strategy is immune.  It requires to navigate safely but
constructively between experiences of reality on one side, and mathematical
abstractions of it, on the other.\smallskip

\noindent We propose to investigate this central difficulty of the modelling
process with a theoretical stance. Thus, considering the second point of view
mentioned above, we do not aim at deriving new mathematical results describing
the properties of models. Conversely, we do not either address the question of
choosing the formal framework to build models for a given category of real
systems.  Instead, we consider problems in which the general lines of the theory
to be used for modelling are assumed to have been set \emph{a priori} and
accepted definitely.  This way, without needing any thorough knowledge on the
systems that are intended to be modelled, we can focus on the \emph{medial} part
of the modelling process, \ie, the part that consists in translating reality
into abstraction and, dually, confronting abstraction to reality within a
predefined formal framework.\smallskip

\noindent In this article, the framework that is considered is that of
\BAN{s}. It serves as a basis to the rest of our discussion. This results in
three notable restrictions to the scope of our arguments.  First, we consider
only the modelling of real systems that are known or supposed to be networks of
interacting elements. Second, we suppose that neither the occurrence of events
nor the mechanisms that are responsible for them can be observed directly. Only
the outcome of events is considered, \ie, only the states (or sequences of
states) in which the network elements end up, as a result of unobserved events,
are effectively observed.  Thirdly, we assume that elements in the networks have
two dual (or possibly extreme) states that can be modelled by theoretical
entities, namely {\em Boolean automata}, that can only take two different
states, $0$ and $1$. In most cases, this last restriction may appear as an
excessive oversimplification of reality.  If the elements in a network do take
more than two states, then, it is probable that the whole range of their
different states and the subtle nuances between them impact appreciably on the
behaviour of other network elements and, \emph{a fortiori}, on the global
network behaviour.  Consequently, the system may satisfy properties that are
likely to elude a ``Boolean modelling'' which can only, by nature, focus on the
roughest and the most obvious events, such as switches between two extreme
states.  For this reason, it may be argued that in some cases, a modelling that
uses \emph{multi-state} automata rather than Boolean automata is better
suited~\cite{Mazoyer1987,Snoussi1993,Thomas1995,Thieffry1995,Richard2007,Richard2009}.
However, we believe that in many cases, precisely because Boolean models can
only hope to produce information on very basic, global properties of a system,
their explanatory scope may be less ambiguous and more easy to exploit than that
of more refined models that account for more subtle and complex properties by
relying on a wider range of parameters.  Thus, although the knowledge that
Boolean modelling helps to develop is much more qualitative than quantitative
and certainly very partial, we believe that it potentially provides reliable,
well-bounded information or, at least, insights on existing causal
relationships, that can serve as solid grounds for further and finer modellings.
As mentioned above, however, our objective here is not to discuss the question
of the pertinence of a general theoretical framework used to represent a part of
reality. Thus, from now on, the Boolean framework is supposed to be accepted and
\BAN{s} are taken to be effective possible satisfactory models of real systems
such as, for instance, sets of genes in a cell, interacting via their protein
products~\cite{Mendoza1998,Demongeot2003,Remy2008a,Demongeot2010}.\smallskip

\noindent In the next section, Section~\ref{SEC-generaldefinitions}, we list the
main characteristics of these networks that are considered in the
literature. This lays the grounds of what we call the ``theory of
\BAN{s}''. Then, in Section~\ref{SEC-theorisation}, we analyse the modelling of
time and causality. This aims at illustrating some problems that arise within
the ``theorisation'' step of the modelling process which consists in completing
the definition of the theory to be used for modelling and simultaneously
specifying a correspondence between the modelled features of a category of real
systems and the modelling features of the theory. The next section,
Section~\ref{SEC-effective-modelling}, deals with the effective modelling of
real systems. It considers the problem of observing the behaviour of a
particular real system (belonging to the category of real systems for which the
theorisation was intended) and deriving from the resulting, necessarily partial
information an understanding of the causes of the events that are observed. Both
Sections~\ref{SEC-theorisation} and~\ref{SEC-effective-modelling}, in the
respective contexts of theorisation and effective modelling, focus on hypotheses
that follow naturally and inevitably from choices involved in the formalisation
of reality and from the necessity to bypass the incompleteness of the
information that is available concerning a certain part of reality. Finally, the
last section of this article, Section~\ref{SEC-discussion}, emphasises the
importance of these hypotheses and addresses the more general issue of the ins
and outs of modelling with Boolean automata networks.

\section{Main features of \BAN{s}}
\label{SEC-generaldefinitions}

Informally, an \emph{automata network of size $n$} involves a set of $n$
multi-state elements interacting with one another. The elements are called
\emph{automata}\footnote{As detailed later, the reason why the term
  \emph{automata} is used here to refer to network elements is that these
  elements are supposed to compute output data, that is, their own state, from
  some given input data, that is, the states of other network
  automata.}~\cite{Robert1995,Robert1986,Choffrut1988,Weisbuch1991}. In the
general case, their set of possible states is any (finite) discrete set. In the
present particular case of \emph{\BAN{s}}, they are supposed to take only two
possible states, $0$ (\emph{inactive}) and $1$ (\emph{active}). The interactions
between automata of a network consist in influences of some automata states on
other automata states. For the sake of simplicity, we abuse language and rather
speak of \emph{influences between automata}.  When or under what circumstances
do these influences effectively produce changes of automata states is a non
trivial question that can be answered several ways according to the purpose of
the study. The present section aims at completing this rough description and
setting up the backbone of the ``theory of \BAN{s}'' by enumerating the key
features of these networks.

\subsection{States and configurations}\label{SEC-States and configurations}

First of all, we introduce some conventions and notations. In the sequel, unless
specified otherwise, the \BAN{s} that are considered are supposed to have size
$n\in\N$ and their automata are assumed to be numbered from $0$ to $n-1$. The
set $V=\{0,\ldots,n-1\}$ refers to the set of network automata. As mentioned
above, automata are supposed to have only two possible states. The binary set
containing these two states is denoted by $\B = \01$. Global states of networks,
called \emph{configurations} in the sequel, are vectors of the set $\Bn$. If $x
= (x_0, \ldots, x_{n-1})\in \Bn$ is the network configuration, then the
\iieme{i} component $x_i\in \B$ of this vector is the \emph{state} of the
\iieme{i} network automaton. In our context, focus is put especially on switches
of automata states starting in a given network configuration. For this reason,
the following notations concerning network configurations will be
useful\footnote{$\uplus$ denotes the disjoint reunion of sets ($A=B\uplus C$
  $\iff$ $\big[ A=B\cup C$ and $B\cap C=\emptyset \big]$) and $\neg$ denotes the
  negation of a Boolean value ($\neg 0=1$ and $\neg 1=0$).}:
\begin{multline}
  \label{barre}
  \forall x = (x_0, \ldots, x_{n-1}) \in \Bn,\ \\
  \begin{array}{l}
    \text{(1)\quad} \forall i\in V,\ \overline{x}^i = (x_0, \ldots, x_{i-1}, 
    \neg x_i, x_{i+1}, \ldots, x_{n-1})\text{,}\\
    \text{(2)\quad} \forall W = W' \uplus \{i\} \subseteq V,\ \overline{x}^W = 
    \overline{\overline{x}^i}^{W'} = \overline{\overline{x}^{W'}}^{i} 
    \text{ and,}\\
    \text{(3)\quad} \overline{x} = \overline{x}^{V} = (\neg x_0, \ldots, \neg 
    x_{n-1})\text{.}
  \end{array}
\end{multline}

\subsection{Structure of a \BAN{}}
\label{SEC-Structure}

To describe a \BAN{}, it is often convenient to start by representing its
underlying \emph{interaction structure} by a digraph $\G=(V,A)$. This digraph is
called the \emph{interaction graph} of the network. Its set of nodes $V$ is
assimilated to the set of automata of the network and its set of arcs $A$
represents the set of interactions that take place in it. More precisely, an arc
$(j,i)\in A$ of this digraph represents the influence that (the state of)
automaton $j\in V$ may possibly have on (the state of) automaton $i\in V$. Let
us note that for the arc $(j,i)$ to belong to $A$, node $j$ is not supposed to
have a \emph{constant} effective impact on $i$. It is merely supposed to have an
impact in \emph{some} network configurations and in at least one of them (see
Equation~(\ref{defarc}) below). In some works, a digraph $\G(x)=(V, A(x))$ is
defined for every configuration
$x\in\Bn$~\cite{Richard2007,Remy2008a,Remy2008b,Siebert2009,Richard2011}. It
contains arcs $(j,i)\in A(x)$ such that in $x$, $j$ does indeed have an
appreciable influence on $i$.  This way, the set of arcs of the interaction
graph $\G$ equals $A=\bigcup_{x\in\Bn}A(x)$.


\begin{example}
  \label{EX-structure-E1}
  The following figure represents an interaction graph $\G=(V,A)$ where
  $V=\{0,1,2\}$ and $A=\{(0,1), (1,1), (1,2), (2,1)\}$. The network whose
  structure it represents thus contains three interacting automata.  Automaton
  $0$ is influenced by no automaton of the network which means that it always
  tends to take the same state. Automaton $1$ is influenced
  by all three of the network automata, including itself. And automaton $2$ is
  influenced just by automaton $1$.\\[2mm] \centerline{
    \begin{minipage}{0.02\textwidth}
      ~~~
    \end{minipage}
    \begin{minipage}{0.95\textwidth}
      \centerline{\scalebox{0.7}{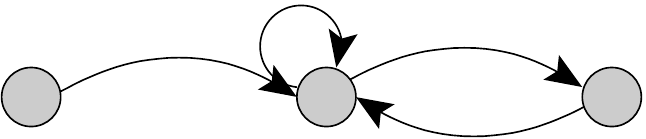}}
    \end{minipage}
    \begin{minipage}{0.02\textwidth}
      \vspace*{4mm}\eex
    \end{minipage}
  }
\end{example}

\subsection{Local transition functions}
\label{SEC-LTF}

The interaction graph of a network gives the existence of the oriented
interactions that it involves. However, it does neither specify the nature of
these interactions nor the conditions under which they effectively occur. This
is done by assigning a \emph{\LTF{}} $f_i:\Bn\to \B$ to each automaton $i\in V$
so that the following is satisfied:
\begin{equation}
  \label{defarc}
  \exists x\in\Bn,\ f_i(x)\neq f_i(\overline{x}^{j})\ \Leftrightarrow\ (j,i) \in
  A\text{.}
\end{equation}

\noindent  With this new definition, the digraphs $\G(x),\, x\in \Bn$ mentioned above contain the
following set of arcs: $A(x)= \{(j,i)\, |\, f_i(x)\neq f_i(\overline{x}^{j})\}$.

\begin{example}
  \label{EX-LTF-E1}
  In agreement with the interaction structure of Example~\ref{EX-structure-E1},
  automata $0$, $1$ and $2$ could, for instance, be assigned the following
  \LTF{s}, respectively:\\[2mm]
  \begin{minipage}{0.02\textwidth}
    ~~~
  \end{minipage}
  \begin{minipage}{0.95\textwidth}
    \begin{equation*}
      \forall x\in \Bn,\ f_0(x) = 1,\ f_1(x) = x_1 \vee (x_0 \wedge \neg x_2),\ 
      f_2(x) = \neg x_1\text{.}
    \end{equation*}
  \end{minipage}
  \begin{minipage}{0.02\textwidth}
    \vspace*{1mm}\eex
  \end{minipage}
\end{example}

\subsection{Events and updates}
\label{SEC-Updates}

In any network configuration, zero, one or several elementary or punctual events
may take place. Here, we consider punctual events that consist in the update of
one or several automata states. We call them respectively atomic and non-atomic
updates and define them formally below.\smallskip

\noindent Supposing that the network is currently in configuration $x\in\Bn$, we
say that automaton $i\in V$ is \emph{updated} if its state switches from $x_i$,
its current state, to $f_i(x)$, its new state. Let us note that, possibly,
$f_i(x)=x_i$ so that the update of $i$ is not effective in $x$. In any case,
this local event yields a global network configuration change (possibly not
effective) which is described by the \emph{$i$-update function} $F_i:\Bn\to\Bn$:
\begin{equation}
  \label{EQ_iupdatefunction}
  \forall x\in \Bn,\ F_i(x) = (x_0,\ldots x_{i-1}, f_i(x), x_{i+1},\ldots, x_{n-1})
  \text{.}
\end{equation}
This event is said to be \emph{atomic} because it involves only one
automaton. We also consider \emph{non-atomic} events that correspond to the
simultaneous update of several automata. In the general case, the
\emph{$W$-update function}\footnote{$\forall i\in V,\, F_i$ obviously equals
  $F_{\{i\}}$ but in this paper, we prefer the first notation.} $F_W:\Bn\to\Bn$
describes the network configuration change that results from the update of all
automata in an arbitrary set $W\subseteq V$:
\begin{equation}
  \label{EQ_Wupdatefunction}
  \forall x\in\Bn,\ \forall i\in V,\ F_W(x)_i = \begin{cases} 
    f_i(x) & \text{if } i\in W\text{,}\\ 
    x_i & \text{otherwise.}
  \end{cases}
\end{equation}
\begin{example}
  \label{EX-updates-E1}
  The following table defines the update functions $F_1$ and $F_{\{0,2\}}$ for
  the network considered in Examples~\ref{EX-structure-E1} 
  and~\ref{EX-LTF-E1}:\\[2mm]
  \begin{minipage}{0.02\textwidth}
    ~~~
  \end{minipage}
  \begin{minipage}{0.95\textwidth}
    \begin{equation*}
      \begin{array}{|c||c|c|c|c|c|} \hline
        x=(x_0,x_1,x_2) & f_0(x) & f_1(x) & f_2(x) & F_1(x) & F_{\{0,2\}}(x)\\
        \hline\hline
        (0,0,0) & 1 & 0 & 1 & (0,0,0) & (1,0,1)\\
        \hline
        (0,0,1) & 1 & 0 & 1 & (0,0,1) & (1,0,1)\\
        \hline
        (0,1,0) & 1 & 1 & 0 & (0,1,0) & (1,1,0)\\
        \hline
        (0,1,1) & 1 & 1 & 0 & (0,1,1) & (1,1,0)\\
        \hline
        (1,0,0) & 1 & 1 & 1 & (1,1,0) & (1,0,1)\\
        \hline
        (1,0,1) & 1 & 0 & 1 & (1,0,1) & (1,0,1)\\
        \hline
        (1,1,0) & 1 & 1 & 0 & (1,1,0) & (1,1,0)\\
        \hline
        (1,1,1) & 1 & 1 & 0 & (1,1,1) & (1,1,0)\\
        \hline
      \end{array}
    \end{equation*}
  \end{minipage}
  \begin{minipage}{0.02\textwidth}
    \vspace*{3.85cm}\eex
  \end{minipage}
\end{example}

\noindent Let us emphasise that the \emph{punctuality} of events mentioned above
refers to their happening in a \emph{unique step} whereas the \emph{atomicity}
of events characterises their \emph{nature}. All atomic as well as all
non-atomic events are punctual. No other punctual events are considered here but
the next paragraph mentions more general events consisting in series of
successive punctual events.

\subsection{Transitions and paths}
\label{SEC-Transitions}

Network \emph{transitions} are couples $(x,y)\in \Bn\times \Bn$ that represent
changes of network configurations (from $x$ to $y$) due to the occurrence of one
or a series of punctual events. Transitions that involve only one punctual
event are called \emph{elementary}. They satisfy $y = F_W(x)$ for some
(possibly empty) set $W\subseteq V$ of automata and are denoted as follows:
\begin{equation*}
  x\trans y,\ x\trans[W]y \text{ or } x\Trans[W]y\text{.}
\end{equation*}
There are two main types of elementary transitions. \emph{Asynchronous
  transitions} correspond to atomic updates. \emph{Synchronous transitions} 
correspond to non-atomic updates. When emphasis needs to be put on the
asynchronicity (resp. synchronicity) of an elementary transition
$x\Trans[\{i\}]y = F_i(x)$ (resp. $x\Trans[W]y = F_W(x)$, $|W|>1$) it is rather
written:
\begin{equation*}
  x\seq y\text{, }x\seq[i] y \text{ or } x\Seq[i] y\text{ (resp. } x\pll 
  y\text{, }x\pll[W]y \text{ or } x\Pll[W]y\text{).}
\end{equation*}
General network \emph{transitions} $(x,y)\in \Bn\times \Bn$ are sequences of
zero, one or several elementary transitions. They are denoted using the
reflexive and transitive closure $\transRT[]$ of $\trans[]$ and defined formally
by:
\begin{multline}
  \label{EQ-transRT}
  x\transRT y \iff\\
  \exists \ell \in \N,\ \exists x^1, \ldots, x^{\ell-1}\in\Bn,\ x\trans x^1\trans
  \ldots\trans x^{\ell-1} \trans y\text{.}
\end{multline}
Any network transition $x\transRT y$ thus corresponds to an ordered list of sets
$(W_t)_{1\leq t\leq \ell}$ such that $ y=F_{W_\ell}\circ \ldots\circ
F_{W_2}\circ F_{W_1}(x)$. When this list is known, we use the following notation
to specify the sequence of punctual updates in question:
\begin{equation*}
  x\transRTW[W_1,W_2,\ldots,W_\ell] y\text{.}
\end{equation*}
Now, for the discussion that follows, it is important to note that because
transitions involve no other events than automata updates, some situations need
to be disregarded. The most basic example is the following:
\begin{equation*}
  x\trans y\text{ where }\exists i\in V,\ x_i = f_i(x) \neq y_i\text{.}
\end{equation*}
Thus, for the network of Examples~\ref{EX-structure-E1} to \ref{EX-updates-E1},
the elementary transition
\begin{equation*}
  (0,0,0)\trans (1,1,0)
\end{equation*} 
is impossible although the network can, all the same, perform the non-elementary
transition
\begin{equation*}
  (0,0,0)\transRT (1,1,0)
\end{equation*} 
by carrying out the sequence
\begin{equation*}
  (0,0,0)\trans (1,0,0) \trans (1,1,0)\text{.}
\end{equation*}
Similarly, suppose that in configuration $x$, two allegedly elementary
transitions, $x\Trans[W] y$ and $x\Trans[W']y'$, are possible. Because any
automaton $i$ that is updated by both transitions necessarily takes state
$f_i(x)$ in both resulting configurations $y$ and $y'$, it must hold that
$\forall i\in W\cap W',\ y_i= y_i'=f_i(x)$. Consequently, the following
situation is also impossible:
\begin{equation*}
  \begin{tabular}{cc}
    \begin{tikzpicture}
      [descr/.style={fill=white,inner sep=2.5pt}]
      \node at (0,1) (x) {$x$}; 
      \node at (2,1.5) (y) {$y$} ;           
      \node at (2,0.5) (z) {$y'$} ;
      \path[->,font=\scriptsize,>=angle 60,] (x) edge[->] node[descr] {$W$} 
      (y);
      \path[->,font=\scriptsize,>=angle 60,] (x) edge[->] node[descr] {$W'$} 
      (z);
    \end{tikzpicture} & 
    \raisebox{0.7cm}{~~where $\exists i\in W\cap W',\ y_i\neq y_i'$.}
  \end{tabular}
\end{equation*}
However, again, a similar situation might be possible if the transitions
$x\transRT y$ and $x\transRT y'$ are not supposed to be elementary. Thus, as
discussed further in Section~\ref{SEC-effective-modelling}, the nature of
transitions observed is an essential precision in the observation of the
behaviour of a network. To determine it, prior knowledge on the network \LTF{s}
is required.\smallskip

\noindent \emph{Paths} (usually called \emph{trajectories} in the context of
\DS{s}) are ordered lists of network transitions $(x^0,x^1), (x^1,x^2),\ldots,
(x^{\ell-1},x^\ell)$, simply written as follows:
\begin{equation*}
  x^0\transRT x^1 \transRT x^2\transRT \ldots \transRT x^{\ell-1} \transRT x^\ell
  \text{.}
\end{equation*}
Since transitions are either elementary or non-elementary, the definition of an
arbitrary path may involve steps corresponding to punctual events as well as
steps corresponding to undetailed \emph{series} of punctual events (see
Example~\ref{EX-traj}). Paths that involve only elementary transitions are said
to be \emph{elementary}.
\begin{example}
  \label{EX-traj}
  The path bellow involves two elementary transitions, $x^0\trans x^1=F_i(x)$
  and $x^2 \trans x^3= F_{W}(x^2)$, as well as one non-elementary transition,
  $x^1 \transRT x^2$, which could itself be broken into a path of several
  elementary transitions if the updates it involves were known:\\[2mm]
  \begin{minipage}{0.02\textwidth}
    ~~~
  \end{minipage}
  \begin{minipage}{0.95\textwidth}
    \begin{equation*}
      x^0\seq[i] x^1\ \transRT\  x^2 \trans[W] x^3 
      \text{.}
    \end{equation*}
  \end{minipage}
  \begin{minipage}{0.02\textwidth}
    \vspace*{2mm}\eex
  \end{minipage}
\end{example}

\subsection{Update schedules}
\label{SEC-us}

An \emph{update schedule} $\delta$ of a set $V$ of automata (or, by extension,
of a network whose set of automata is $V$), is defined by an ordered (finite or
infinite) list $(W_t)_{t\in S}$ ($S\subseteq \N$) of non-empty sets of automata
($\forall t\in S,\ \emptyset\neq W_t\subseteq V$). We write $\delta \equiv
(W_t)_{t\in S}$ or just $\delta \equiv W_0, W_1, \ldots, W_t, \ldots$.  Under an
update schedule $\delta\equiv (W_t)_{t\in S}$, starting in configuration
$x\in\Bn$, a network takes sequentially the configurations $x^0 = F_{W_0}(x)$,
$x^1 = F_{W_1}\circ F_{W_0}(x)$, $\ldots$, $x^t = F_{W_t}\circ \ldots \circ
F_{W_0}(x)$, $\ldots$, \ie, it follows the elementary path:
\begin{equation} 
  \label{EQ-US-traj}
  \begin{tikzpicture}[descr/.style={fill=white,inner sep=2.5pt}]
    \node at (0,0) (x) {$x$}; 
    \node at (1.7,0) (x0) {~$F_{W_0}(x)$~} ;           
    \node at (3.6,0) (x1) {~$\ldots$~} ;
    \node at (6.35,0) (xt) {~$F_{W_t}\circ \ldots \circ  F_{W_0}(x)$~} ;
    \node at (9.3,0) (xz) {~$\ldots$~} ;
    \path[->,font=\scriptsize,>=angle 60,] (x) edge[->] node[above] {$W_0$} 
    (x0);
    \path[->,font=\scriptsize,>=angle 60,] (x0) edge[->] node[above] {$W_1$} 
    (x1);
    \path[->,font=\scriptsize,>=angle 60,] (x1) edge[->] node[above] {$W_t$} 
    (xt);
    \path[->,font=\scriptsize,>=angle 60,] (xt) edge[->] node[above] {$W_{t+1}$} 
    (xz);
  \end{tikzpicture}
\end{equation}
In particular, $\delta$ only allows the network to perform elementary
transitions that update one of the sets $W_t$, $t\in S$. If $U\subseteq V$ is a
set of automata that differs from all of these sets ($U\neq W_t$, $\forall t\in
S$), then $x\Trans[U] F_U(x)$ is not an elementary transition that can be done
under $\delta$. In addition, the sets $W_t$, $t\in S$ themselves cannot either
be updated in any network configuration. To detail this, let us introduce by
induction on $t\in S$ the sets $X_t = \{x\ |\ x\Trans[W_t] F_{W_t}(x)$ is
allowed by $\delta\}\subseteq\Bn$:
\begin{equation}
  \label{EQ-Xt}
  \left\{
    \begin{array}{l}
      X_0 = \Bn\text{,}\\ 
      \forall t\in S,\ X_{t+1} = \{F_{W_t}(x)\ |\ x\in X_t\}\, =\, F_{W_t}(X_t)\text{.}
    \end{array}
  \right.
\end{equation}
Example~\ref{EX-US-E1} below illustrates that for any configuration $x\notin
X_t$, by definition of $X_t$, the elementary transition $x\Trans[W_{t}]
F_{W_{t}}(x)$ is not possible according to $\delta$.
\begin{example}
  \label{EX-US-E1}
  Suppose that the network considered in Examples~\ref{EX-structure-E1}
  to~\ref{EX-updates-E1} is updated by the periodic update schedule
  $\delta\equiv \{1\},\{0,2\},\{1\},\{0,2\},\ldots$. Then, since $\delta$ does
  not allow the {\em atomic} update of automaton $2$, any elementary transition of the
  form $(0,x_1,x_2)\Trans[\{2\}] (0,x_1,\neg x_1)$ is impossible. The sets
  $X_t,\ t\in\N$ defined above can be shown to equal the following: $X_0=\B^3$,
  $ X_1=\B^3 \setminus \{(1,0,0)\}$ and, $\forall t\geq 2$,
  $X_t=\{(1,0,1),(1,1,0)\}$. As a consequence, in configuration $(1,0,0)$,
  $\delta$ does not either allow the set $\{0,2\}$ to be updated.\eex
\end{example}


\noindent As mentioned in Section~\ref{SEC-Transitions}, our choice of
  definitions imposes that certain network behaviours be baned.  Adding the
  supplementary constraint of an update schedule restricts further the
  situations that may be considered possible. For instance, the following
  situation in which $x$ belongs to both the sets $X_3$ and $X_5$ (see
  Equation~\ref{EQ-Xt} above) is consistent with the \us{} $\delta$ only if $\forall
  i\in W_3\cap W_5$ it holds that $y_i=z_i=f_i(x)$:\\
\centerline{
  \begin{tikzpicture}[descr/.style={fill=white,inner sep=2.5pt}]
    \node at (-6,0.9) (y0) {~$y^0$~}; 
    \node at (-4,0.65) (y1) {~$y^1$~}; 
    \node at (-2,0.4) (y2) {~$y^2$~}; 
    \node at (1.1,0) (x) {$\begin{array}{cc}x=F_{W_2}(y^2)\\=F_{W_4}(z^4)\end{array}$~};
    \path[->,font=\scriptsize, >=angle 60,] (y0) edge[->] node[above] {$W_0$} (y1); 
    \path[->,font=\scriptsize, >=angle 60,] (y1) edge[->] node[above] {$W_1$} (y2);   
    \path[->,font=\scriptsize, >=angle 60,] (y2) edge[->] node[above] {$W_2$} (x);
    \node at (-4,-0.6) (z3) {~$z^3$~}; 
    \node at (-2,-0.4) (z4) {~$z^4$~}; 
    \draw[-,dotted,thick,] (-6.6,-0.96)--(-6,-0.9);
    \path[->,font=\scriptsize, >=angle 60,] (-5.8,-0.85) edge[->] node[above] {$W_2$} (z3); 
    \path[->,font=\scriptsize, >=angle 60,] (z3) edge[->] node[above] {$W_3$} (z4);   
    \path[->,font=\scriptsize, >=angle 60,] (z4) edge[->] node[below] {$W_4$} (x);
    \node at (5.2,-0.45) (y) {~$y=F_{W_3}(x)$~}; 
    \node at (5.2,0.45) (z) {~$z=F_{W_5}(x)$~}; 
    \path[->,font=\scriptsize, >=angle 60,] (x) edge[->] node[above] {$W_5$} (z);
    \path[->,font=\scriptsize, >=angle 60,] (x) edge[->] node[below] {$W_3$} (y); 
\end{tikzpicture}}

 \noindent and generally, for any subsets $W_t, W_{t'}\subseteq V$ that belong to the
  defining list of an \us{} $\delta$, the following must hold:\\ \centerline{$\forall
    x\in \Bn$, $\big[y=F_{W_t}(x)$ and $z=F_{W_{t'}}(x)\big]$ $\Rar$
    $\big[\forall i\in W_t\cap W_{t'},\ y_i=z_i=f_i(x)\big]$.} Thus, again, the
  set of \LTF{s} of a network updated with a given \us{} needs to be known in
  order to determine the trajectories that are possible.  \smallskip


\noindent \emph{Periodic update schedules} of arbitrary period $p\in\N$
correspond to infinite periodic lists $W_0, W_1, \ldots, W_{p-1},W_0, W_1,
\ldots, W_{p-1},\ldots $ (\eg, the update schedule of period $2$ in
Example~\ref{EX-US-E1}). For the sake of simplicity they are rather defined by
finite ordered lists $(W_t)_{t\in \NN[p]}$ of size $p$: $\delta\equiv
W_0,W_1,\ldots,W_{p-1}$. We define \emph{global transition functions}
$F[\delta]:\Bn\to \Bn$ relative to such update schedules:
\begin{equation}
  \label{EQ-GTF}
  \forall x\in\Bn,\ F[\delta](x) = F_{W_{p-1}}\circ \ldots \circ F_{W_1}\circ
  F_{W_0} (x)\text{.}
\end{equation}
The definition of this function allows to focus on series of $p$ elementary
transitions rather than on single elementary transitions so that
Equation~\ref{EQ-US-traj} can be simplified to the following (non necessarily
elementary) path: 
\begin{equation*}
  x\transRT F[\delta](x)\transRT F[\delta]^2(x)\transRT \ldots \transRT 
  F[\delta]^k(x)\transRT\ldots\text{,} 
\end{equation*}
where $F[\delta]^k$ denotes the \iieme{k} iterate of $F[\delta]$.\smallskip

\noindent Now, let $\approx$ be the equivalence relation that relates periodic
\us{s} that differ only by a rotation of their sequence of updates. For
instance, under this relation, the periodic \us{} defined by the list $\{1\},
\{0,2\}, \{1,2\}$ is equivalent to the two periodic \us{s} defined by the lists
$ \{0,2\}, \{1,2\}, \{1\}$ and $\{1,2\}, \{1\}, \{0,2\}$.  Consider two
equivalent \us{s} $\delta\equiv W_0,W_1,\ldots,$ $W_{p-1} $ and $\delta'\equiv
W'_0,W'_1,\ldots,$ $W'_{p-1} $ satisfying $\forall t\in
\NN[p],\ W_t'=W_{t+\Delta}$ for some $\Delta\in \NN[p]$. Then, let us note that
any elementary path starting in $x\in \Bn$ under $\delta$ becomes identical, at
its \iieme{\Delta} step, to the elementary path that starts in $F_{W_{\Delta
    -1}}\circ\ldots\circ F_{W_0}(x)$ under $\delta'$. Thus, except for $\Delta$
elementary transitions at the beginning of each path, $\delta$ and $\delta'$
yield exactly the same network behaviours. Focusing on non-elementary
transitions representing series of $p$ elementary transitions, however, the two
\us{s} yield very different paths that cannot generally be identified at all
from the point of view of an outside observer that knows nothing about the
update schedule that is used.\smallskip

\label{USs as fcns}
\noindent Let us also note that \PERus{s} $\delta$ may also be defined as
functions $\delta : V\to {\cal P}(\NN[p]))$, where ${\cal P}(S)$ refers to the
power set of a set $S$ (see Example~\ref{EX-US} below). This way, for any
automaton $i\in V$, $\delta(i)$ is the set of updates involving node $i$ in the
periodic sequence $(W_t)_{t\in \NN[p]}$ such that:
\begin{equation*}
  \forall t\in \NN[p],\ i\in W_t \iff t\in \delta(i)\text{.}
\end{equation*}
And since each subset $W_t$, with $t\in \NN[p]$, must be non-empty in order for
the \us{} to effectively have  period $p$, $\delta$ must satisfy:
\begin{equation*}
  \forall t\in \NN[p],\ \exists i\in V,\ t\in \delta(i)\text{.}
\end{equation*}
A well-known instance of \PERus{s} are \emph{block-sequential
  \us{s}}~\cite{Robert1995,Robert1986,Elena2004,Tournier2005,Demongeot2008a,Aracena2009,Goles2010}.
Their particularity lies in that their sequence of updates involves exactly once
each automaton of the network. Thus, they can be defined either by a finite list
$(W_t)_{t\in \NN[p]}$ such that $V = \biguplus_{{t\in \NN[p]}} W_t$ or, abusing
notations introduced above, by a function $\delta : V\to \NN[p]$ (see
Example~\ref{EX-US}). The \emph{parallel update schedule} is the unique \BSus{}
of period $p = 1$ ($\forall i\in V,\ \delta(i) = 1$). It updates all nodes of
the network in one step, simultaneously. \emph{Sequential \us{s}} are \BSus{s}
with period equal to the size of the network ($p = n$). They update only one
node of the network at a time ($\forall t\in \NN[p],\ |W_t| = 1$). Let us
mention here that identifying the \BSus{s} that are equivalent under $\approx$
reduces the number of these \us{s}
\footnote{Proof. Let $S(n,k)$ for $k\leq n$ count the number of surjective
  applications from a set of $n$ elements to a set of $k$. The number of
  \BSus{s} of a set of $n$ automata equals $\BSn = \sum_{0\leq
    k<n}\binom{n}{k}\cdot\BSk[k]\ =\ \sum_{1\leq k \leq n} S(n,k)\sim
  \frac{1}{2}\cdot \frac{n!}{(ln\ 2)^{n+1}}$~\cite{Wilf1990} (sequnce A670
  in~\cite{Sloane2002}). The number of equivalence classes for the relation
  $\approx$ can be shown to equal $\BSnH = \sum_{1\leq k \leq
    n}\frac{S(n,k)}{k}$.  Thus, from $S(n+1,k)=k(S(n,k)+S(n,k-1)$, results that
  $ \BSkH[n+1]= 2\times\BSn $ and $ \BSnH \sim \frac{2\cdot
    ln\ 2}{n}\times\BSn$.} by a factor that tends towards $\frac{2\cdot
  ln\ 2}{n}$.\smallskip

\noindent A larger class of \PERus{s} that contains all \BSus{s} is the class of
\emph{\Sus{s}} (mentioned later in Section~\ref{SEC-infer} for their
particularity). This class contains all \PERus{s} that do not update any
automata more than once within each period. With the functional notation
introduced earlier on Page~\pageref{USs as fcns}, \Sus{s} are defined as the
\PERus{s} satisfying:
\begin{equation}
  \forall i\in V,\ |\delta(i)|\leq 1\text{.}
\end{equation}

\noindent Finally, let us introduce another generalisation of \BSus{s}.
\emph{Fair \us{s}}~\cite{Tosic2006,Goles2011} are the \PERus{s} that update
each automaton at least once (see Example~\ref{EX-US}). Unlike \BSus{s}, they
may update some automata more often than others. A $k$-fair \us{} of period
$p$ is an \us{} $\delta : V\to {\cal P}(\NN[p]))$ such that for all automata $i$
and $j$ of the network, the following holds:
\begin{equation*}
  |\delta(i)|\leq k\cdot |\delta(j)|\text{,}
\end{equation*}
\ie, within each period, $i$ is not updated more than $k$ times as much as $j$
is. Block-sequential \us{s} are a special type of $1$-fair \us{}.
\begin{example}
  \label{EX-US}
  Consider a network of size $n = 6$. The $3$-fair \us{} $\delta\equiv
  \{2,5\}\{0,1,4\}\{1,2,3\}\{0,1,4,5\}$, the block-sequential \us{} $\beta$
  $\equiv$ $\{2\}$ $\{3,4\}$$\{0,1,5\}$, the sequential \us{} \mbox{$\sigma \equiv
    \{5\}\{3\}\{1\}\{0\}\{2\}\{4\}$} and the parallel \us{} $\pi\equiv
  \{0,1,2,3,4,5\}$ can be defined as functions:\\[2mm]
  \begin{minipage}{0.97\textwidth}
    \begin{center}
      \hspace*{1.2cm}\begin{tabular}{m{5cm}m{1cm}m{5cm}}
        $\delta:\left\{\begin{array}{ccc}
            V & \to & {\cal P}(\NN[4])\\
            0 & \mapsto &\{1,3\}\\
            1 & \mapsto &\{1,2,3\}\\
            2 & \mapsto &\{0,2\}\\
            3 & \mapsto &\{2\}\\
            4 & \mapsto &\{1,3\}\\
            5 & \mapsto &\{0,3\}
          \end{array}\right.\text{,}$
        & ~\qquad~ & 
        \vspace*{9mm}$\sigma:\left\{\begin{array}{ccc}
            V & \to &  \NN[6]\\
            0 & \mapsto & 3\\
            1 & \mapsto & 2\\
            2 & \mapsto & 4\\
            3 & \mapsto & 1\\
            4 & \mapsto & 5\\
            5 & \mapsto & 0
          \end{array}\right.\text{,}$\vspace*{3mm}
        \\
        $\beta:\left\{\begin{array}{ccc}
            V & \to &  \NN[3]\\
            i\in\{0,1,5\} & \mapsto & 2 \\
            2 & \mapsto & 0\\
            i\in\{3,4\} & \mapsto & 1
          \end{array}\right.\text{,}$
        & ~\qquad~ & 
        $\pi:\left\{\begin{array}{ccc} V & \to  & \NN[1]\\ 
            \forall i\in V & \mapsto & 0
          \end{array}\right.\text{.}$
      \end{tabular}
    \end{center}
  \end{minipage}
  \begin{minipage}{0.02\textwidth}
    \vspace*{5.6cm}\eex
  \end{minipage}
\end{example}

\subsection{Automata stability and transition effectiveness}
\label{SEC-stability}

As mentioned above, when an automaton is updated in a given configuration
$x\in\Bn$, it does not necessarily change states. We define the set $\U(x)$ of
automata that can indeed change states in $x$ and that do if and only if they
are updated:
\begin{equation*}
  \U(x)=\{i\in V\ |\ f_i(x)\neq x_i\}\text{.}
\end{equation*} 
The automata in $\U(x)$ are said to be \emph{unstable} in $x$ and those in
$\nU(x) = V\setminus \U(x)$ are said to be \emph{stable} in $x$.  It is
important to note that the only couples $(x,y)\in \Bn\times\Bn$ that indeed are
elementary network transitions are the couples that satisfy:
\begin{equation*}
  D(x,y)=\{i\in V\ |\ x_i\neq y_i\}\ \subseteq\ \U(x)\text{.}
\end{equation*} 
It also is important to note that for any set $W\subseteq V$, the following
holds:
\begin{equation*}
  \forall x\in \Bn,\ F_W(x)\ =\ F_{W\cap\, \U(x)}(x)\ =\ 
  \overline{x}^{W\cap\U(x)}\text{.} 
\end{equation*} 
As a consequence, the elementary transitions
\begin{equation*}
  x \trans[W] F_W(x) \text{ and } x \trans[W\cap\, \U(x)] F_{W\cap\, \U(x)}(x) 
\end{equation*} 
perform different updates but produce identical effects. Elementary transitions
$x \Trans[W] y$ such that $W\cap\U(x) = \emptyset$ and $x = y$ are called
\emph{null transitions}. Elementary transitions $x \Trans[W] y$ such that
$W\subseteq \U(x)$ and $y = \overline{x}^{W}$ are called \emph{effective
  transitions}. Other elementary transitions are said to be partially null
or partially effective.

\subsection{State transition systems and transition graphs}
\label{SEC-GT}

\def\smallg{{\text{\footnotesize{\sc g}}}}
\def\smalla{{\text{\footnotesize{\sc a}}}}

A \emph{\TG{}} of a network is a digraph $\T{} = (\States,T)$ that represents
its {\em global behaviour}. Its nodes are network configurations and its arcs
are network transitions. There are several important characteristics that a
\TG{} may have. A first characteristic concerns the nature of the transitions
contained in the digraph.  If all arcs are elementary transitions, then the \TG{}
it is said to be \emph{elementary}.\smallskip

\noindent Among the set of elementary \TG{s} that may be associated to a network
of size $n$ is its \emph{general \TG{}}~\cite{Noual2011a}, \GIG{} for short, denoted by
$\T{\smallg}=(\Bn, T_\smallg)$.  Its set of nodes is $\Bn$
and its set of arcs is the set of {\em all} elementary network transitions:
\begin{equation*}
  T_\smallg=\bigcup\{(x,F_W(x))\ |\ x\in \Bn,\ W\neq \emptyset\,
  \subseteq V\}\text{.}
\end{equation*} 
Every node $x\in\Bn$ in this multiple digraph has out-degree $deg^+(x) =
2^n-1$. Another notable elementary \TG{} is the \emph{asynchronous
  \TG{}}, or
\SIG{} for short. It is is the spanning sub-graph $\T{\smalla}=(\Bn,T_\smalla)$
of $\T{\smallg}$ whose set of transitions equals the set of
asynchronous transitions of the network:
\begin{equation*}
  T_\smalla = \bigcup\{(x,F_i(x))\ |\ x\in \Bn,\ i\in V\}\text{.} 
\end{equation*} 
and in which each node $x\in \Bn$ has out degree $\deg^+(x) = n$ (the size of
the network).  Arbitrary transition graphs that contain only asynchronous
transitions are also called
asynchronous~\cite{Snoussi1993,Remy2008b,Siebert2009,Thomas1991,Harvey1997,Bernot1997,Bernot2004,Richard2010,Delaplace2010}.
Because updates are not always effective, generally, an elementary \TG{} \T{} is
a multiple digraph. Its simple version is called its \emph{effective version}
and is denoted by $\eff{\T{}}$.  In this digraph, all arcs $(x,y),\ x\neq y$
which are not loops are effective transitions, \ie, $D(x,y)\subseteq \U(x)$ (see
Example~\ref{EX-TG-E1} below). The effective version of the \GIG{} is refered to
as the \GGIG{}, and that of the \SIG{} as the \SSIG{}.\smallskip

\noindent \emph{Non-elementary \TG{s}} which contain non-elementary transitions
are often used to describe the behaviour of a network that is updated according
to a certain \PERus{} (see Example~\ref{EX-TG-E1}). To a \PERus{} $\delta\equiv
(W_t)_{t\in \NN[p]}$, we associate the digraph $T_\delta $ of $F[\delta]$ (see
Equation~\ref{EQ-GTF}):
\begin{equation}
  \label{EQ-TGus}
  \T{\delta} = (\Bn,T_\delta) \text{ where } T_\delta = \{(x,F[\delta](x))\ |\ 
  x\in \Bn\}\text{.}
\end{equation}\smallskip

\noindent Let us highlight that a \TG{} $\T{}=(\States,T)$ does not necessarily
have a node set $\States$ included in $\Bn$.  Indeed, when the network behaviour
is \emph{context-dependent}, \ie, when the transitions that are possible in a
given configuration depend on the paths that lead to that configuration, in
order to avoid loosing any information, it may be required to define $\States$
as a {\em multi-set} on $\Bn$. This is the case for {\em elementary} \TG{s} that
describe the behaviour of networks that are updated with \PERus{s} (see
$\T{\delta}^{{\text{\tiny\textsc{elem}}}}$ in Example~\ref{EX-TG-E1}): if
configuration $x$ belongs to $k$ different sets $X_t$ (see
Equation~\ref{EQ-Xt}), then, because the network behaviour is deterministic, any
elementary \TG{} that represents exhaustively the network behaviour needs to
contain $k$ copies of $x$ (consider merging nodes representing the same
configuration in the digraph $\T{\delta}^{{\text{\tiny\textsc{elem}}}}$ of
Example~\ref{EX-TG-E1}). Thus, to a \PERus{} $\delta\equiv (W_t)_{t\in \NN[p]}$,
we also associate the ``elementary version'' of $\T{\delta}$, namely the
elementary graph $\T{\delta}^{{\text{\tiny\textsc{elem}}}}$ which contains all
elementary paths of the network updated with $\delta$ .  Its set of arcs equals
set of transitions $\{(x,F_{W_t}(x))\in X_t\times X_{t+1}\}$.  \smallskip

\noindent When, on the contrary, the network behaviour at one step is
independent of its behaviour at all previous steps, we say that it is
\emph{context-free} (or \emph{memory-less}). In this special case, the \TG{}
$\T{} = (\States,T)$ defines exactly a
\emph{\STS{}}~\cite{Lam1990,Clarke1996,Geurts1998}, that is, a digraph whose set
of nodes $\States\subseteq \Bn$ is the set of states of the system and whose set
of arcs $T\subseteq \States\times \States$ is its set of transitions. Contrary
to $\T{\delta}^{{\text{\tiny\textsc{elem}}}}$, the \TG{s} $\T{\delta}$,
$\T{\smallg}$, $\eff{\T{\smallg}}$, $\T{\smalla}$ and $\eff{\T{\smalla}}$
mentioned above as well as any of their sub-graphs all are examples of \STS{s}.
\begin{example} 
  \label{EX-TG-E1}
  Consider the network of Examples~\ref{EX-structure-E1}
  to~\ref{EX-updates-E1}. Its \GGIG{} is the digraph represented below where the
  set of arcs represented with thicker lines equals the set of arcs of the
  \SSIG{} of the same network.\\[2mm]
  \centerline{
    \begin{tikzpicture}[descr/.style={fill=white,inner
        sep=2.5pt}][descr/.style={fill=white,inner sep=2.5pt}]
      \node at (0,0)   (x0) {$(0,0,0)$~};
      \node at (0,2)   (x1) {$(0,0,1)$~};
      \node at (9,0)   (x2) {$(0,1,0)$~}; 
      \node at (9,2)   (x3) {$(0,1,1)$~}; 
      \node at (3,0)   (x4) {$(1,0,0)$~}; 
      \node at (3,2)   (x5) {$(1,0,1)$~}; 
      \node at (6,0)   (x6) {$(1,1,0)$~}; 
      \node at (6,2)   (x7) {$(1,1,1)$~}; 
      \path[->,font=\scriptsize, >=angle 60,] (x0) edge[->, very thick] 
      node[descr] {$\{2\}$}
      (x1) edge[->] node[descr] {$\{0,2\}$} (x5) edge[->, very thick] 
      node[descr] {$\{0\}$} (x4);
      \path[->,font=\scriptsize, >=angle 60,] (x1) edge[->, very thick] 
      node[descr] {$\{0\}$} (x5);
      \path[->,font=\scriptsize, >=angle 60,] (x4) edge[->, very thick] 
      node[descr] {$\{2\}$} (x5) edge[->] node[descr] {$\{1,2\}$} (x7) 
      edge[->, very thick] node[descr] {$\{1\}$} (x6);
      \path[->,font=\scriptsize, >=angle 60,] (x7) edge[->, very thick] 
      node[descr] {$\{2\}$} (x6);
      \path[->,font=\scriptsize, >=angle 60,] (x2) edge[->, very thick] 
      node[descr] {$\{0\}$} (x6);
      \path[->,font=\scriptsize, >=angle 60,] (x3) edge[->, very thick] 
      node[descr] {$\{0\}$} (x7) edge[->] node[descr] {$\{0,2\}$} (x6) 
      edge[->, very thick] node[descr] {$\{2\}$} (x2);
      \draw (x0) to [->,font=\scriptsize, min distance=10mm, in =-135, out=-45,
      very thick] node[descr] {$\{1\}$} (x0);
      \draw (x1) to [->,font=\scriptsize, min distance=10mm, in =45, out=135, 
      very thick] node[descr] {$\{1,2\}$} (x1);
      \draw (x2) to [->,font=\scriptsize, min distance=10mm, in =-135, out=-45,
      very thick] node[descr] {$\{1,2\}$} (x2);
      \draw (x3) to [->,font=\scriptsize, min distance=10mm, in =45, out=135, 
      very thick] node[descr] {$\{1\}$} (x3);
      \draw (x4) to [->,font=\scriptsize, min distance=10mm, in =-135, out=-45,
      very thick] node[descr] {$\{0\}$} (x4);
      \draw (x5) to [->,font=\scriptsize, min distance=10mm, in =45, out=135, 
      very thick] node[descr] {$\{0,1,2\}$} (x5);
      \draw (x6) to [->,font=\scriptsize, min distance=10mm, in =-135, out=-45,
      very thick] node[descr] {$\{0,1,2\}$} (x6);
      \draw (x7) to [->,font=\scriptsize, min distance=10mm, in =45, out=135, 
      very thick] node[descr] {$\{0,1\}$} (x7);
    \end{tikzpicture}
  } 
  The \TG{} $\T{\delta}$ of the network of Examples~\ref{EX-structure-E1}
  to~\ref{EX-updates-E1} associated to the \us{} given in Example~\ref{EX-US-E1}
  is the following:\\[2mm]
  \centerline{
    \begin{tikzpicture}[descr/.style={fill=white,inner sep=2.5pt}]
      \node at (0,0.5)    (x0) {$(0,0,0)$~};
      \node at (0,0)      (x1) {$(0,0,1)$~};
      \node at (2.5,0.25) (x5) {$(1,0,1)$};
      \node at (5,0.5)    (x2) {$(0,1,0)$~};
      \node at (5,0)      (x3) {$(0,1,1)$~};
      \node at (10,0.5)   (x4) {$(1,0,0)$~};
      \node at (10,0)     (x7) {$(1,1,1)$~};
      \node at (7.5,0.25) (x6) {$(1,1,0)$};
      \path[->,font=\scriptsize, >=angle 60,] (x5) edge[<<-] (x0) edge[<<-]  
      (x1);
      \path[->,font=\scriptsize, >=angle 60,] (x6) edge[<<-] (x2) edge[<<-]  
      (x3) edge[<<-] (x4) edge[<<-] (x7);
    \end{tikzpicture}
  }
  and the corresponding elementary \TG{}
  $\T{\delta}^{{\text{\tiny\textsc{elem}}}}$ is: \\[2mm] \centerline{
    \begin{minipage}{\textwidth}
      \begin{minipage}{0.02\textwidth}
        ~~~
      \end{minipage}
      \begin{minipage}{0.95\textwidth}
        \begin{center}
          \begin{tikzpicture}[descr/.style={fill=white,inner sep=2.5pt}]
            \node at (0,1.5)   (x0) {$(0,0,0)$~}; 
            \node at (2.5,1.5) (x01) {$(0,0,0)$~};
            \node at (5,1.5)   (x02) {$(1,0,1)$}; 
            \node at (0,1)     (x1) {$(0,0,1)$~}; 
            \node at (2.5,1)   (x11) {$(0,0,1)$~};
            \node at (5,1)     (x12) {$(1,0,1)$}; 
            \node at (0,0.5)   (x2) {$(0,1,0)$~}; 
            \node at (2.5,0.5) (x21) {$(0,1,0)$~}; 
            \node at (5,0.5)   (x22) {$(1,1,0)$}; 
            \node at (0,0)     (x3) {$(0,1,1)$~}; 
            \node at (2.5,0)   (x31) {$(0,1,1)$~}; 
            \node at (5,0)     (x32) {$(1,1,0)$}; 
            \path[->,font=\scriptsize, >=angle 60,] (x0) edge[->] node[above] 
            {$\{1\}$} (x01);
            \path[->,font=\scriptsize, >=angle 60,] (x01) edge[->] node[above] 
            {$\{0,2\}$} (x02);
            \path[->,font=\scriptsize, >=angle 60,] (x1) edge[->]  (x11);
            \path[->,font=\scriptsize, >=angle 60,] (x11) edge[->] (x12);
            \path[->,font=\scriptsize, >=angle 60,] (x2) edge[->]  (x21);
            \path[->,font=\scriptsize, >=angle 60,] (x21) edge[->] (x22);
            \path[->,font=\scriptsize, >=angle 60,] (x3) edge[->]  (x31);
            \path[->,font=\scriptsize, >=angle 60,] (x31) edge[->] (x32);
          \end{tikzpicture}\\[3mm]
          \begin{tikzpicture}[descr/.style={fill=white,inner sep=2.5pt}]
            \node at (7,1.5)   (x4) {$(1,0,0)$~}; 
            \node at (9.5,1.5) (x41) {$(1,1,0)$~}; 
            \node at (12,1.5)  (x42) {$(1,1,0)$}; 
            \node at (7,1)     (x5) {$(1,0,1)$~}; 
            \node at (9.5,1)   (x51) {$(1,0,1)$~}; 
            \node at (12,1)    (x52) {$(1,0,1)$}; 
            \node at (7,0.5)   (x6) {$(1,1,0)$~}; 
            \node at (9.5,0.5) (x61) {$(1,1,0)$~}; 
            \node at (12,0.5)  (x62) {$(1,1,0)$}; 
            \node at (7,0)     (x7) {$(1,1,1)$~}; 
            \node at (9.5,0)   (x71) {$(1,1,1)$~}; 
            \node at (12,0)    (x72) {$(1,1,0)$}; 
            \path[->,font=\scriptsize, >=angle 60,] (x4) edge[->] node[above] 
            {$\{1\}$} (x41);
            \path[->,font=\scriptsize, >=angle 60,] (x41) edge[->] node[above] 
            {$\{0,2\}$} (x42);
            \path[->,font=\scriptsize, >=angle 60,] (x5) edge[->]  (x51);
            \path[->,font=\scriptsize, >=angle 60,] (x51) edge[->] (x52);
            \path[->,font=\scriptsize, >=angle 60,] (x6) edge[->]  (x61);
            \path[->,font=\scriptsize, >=angle 60,] (x61) edge[->] (x62);
            \path[->,font=\scriptsize, >=angle 60,] (x7) edge[->]  (x71);
            \path[->,font=\scriptsize, >=angle 60,] (x71) edge[->] (x72);
          \end{tikzpicture}
        \end{center}
      \end{minipage}
      \begin{minipage}{0.01\textwidth}
        \vspace*{4.1cm}\eex
      \end{minipage}
    \end{minipage}
  }
\end{example}

\subsection{Network behaviours}
\label{SEC-behaviours}



If $\T{}= (\States,T)$ is the \TG{} representing the {\em global} behaviour of a
given network, then, any sub-graph of \T{} represents a {\em particular}
behaviour of that network. For instance, if a network is supposed to be
potentially able of performing {\em any} elementary transition, then, \T{}
either equals the \GIG{} $\T{\smallg}$ of the network or its \GGIG{}
$\eff{\T{\smallg}}$ and any elementary \TG{} represents a possible particular
behaviour of the network.\smallskip


\noindent Let us focus on \STS{s} $\T{} = (\Bn,T)$. For these, we define the
binary relation $T^{\ast}\subseteq \transRT$ as the reflexive and transitive
closure of the relation $T$, that is, $(x,y)\in T^{\ast} $ if and only if there
exists in \T{} a path from $x$ to $y$. \emph{Transient configurations} are then
defined as the configurations $x\in\Bn$ that satisfy:
\begin{equation*}
  \exists y\in \Bn,\ (x,y)\in T^{\ast} \wedge (y,x)\notin T^{\ast}\text{.} 
\end{equation*}
Any behaviour of the network that involves transient configurations is said to
be transient itself. Configurations that are not transient are called
\emph{recurrent}. These configurations are precisely those that induce the
terminal strongly connected components of \T{}, called \emph{limit
  behaviours} here (and rather called  \emph{attractors} in the context of
deterministic \DS{s}~\cite{Cosnard1985,Diner1987,Kauffman1993}). There are two
main types of limit behaviours.  Those that contain strictly more than one
configuration are called (sustained) \emph{oscillations}~\cite{Thomas1981} (or
\emph{limit cycles} when \T{} defines a deterministic \DS{}). Limit behaviours
of size one are called \emph{stable configurations} (or \emph{fixed points} when
\T{} defines a deterministic \DS{}\footnote{In this case, \T{} is the graph of a
  global transition function $F:\Bn\to\Bn$ so any stable configuration in this
  graph is a fixed point of that function.}). They are characterised by their
out-going effective degrees\footnote{In any \TG{} \T{}, the {\em out-going
    effective degree} of configuration $x$ is the number of arcs out-going $x$
  that are {\em effective} transitions.} being equal to $0$. Note that in a
stable configuration $x$, it is not necessary that all automata of the network
be stable (\ie, that $\U(x) = \{i\in V\ |\ f_i(x)\neq x_i\}$ be empty). Indeed,
for $x$ to be stable, it suffices that \T{} contains no transition that
corresponds to the update in $x$ of automata belonging to $\U(x)$ (this happens
in particular when $ \T{} =\T{\delta}$).

\subsection{Relationships between network features} 
\label{SEC-infer}

In the sequel, we informally call {\em observer of a network} ``anyone'' that
has full, partial or no knowledge of its characteristics (structure, set of
interactions, behaviour\ldots). In this section, we discuss how the different
features of the network relate and how an observer carrying one type of
information on the network may derive additional information of another
type.\smallskip

\noindent First, the information carried by the interaction graph (as defined in
Sections~\ref{SEC-Structure} and~\ref{SEC-LTF}) of a network is contained in the
information carried by its set of \LTF{s}. However, it takes exponential time in
the size $n$ of the network to draw an interaction graph from a set of
\LTF{s}. This complexity stands even when the \LTF{s} are given in conjunctive
normal form (CNF). Indeed, given $j\in V$ and the CNF definition of $f_i$, the
problem of determining whether there exists $x\in\Bn$ such that $f_i(x)\neq
f_i(\overline{x}^{j})$ is NP-complete: the CNF version of
SAT~\cite{Cook1971,Garey1979} which is NP-complete can be reduced to
it.\smallskip

\noindent Next, let us emphasise that with the sole knowledge of the set of
\LTF{s} of a network, many different \TG{s} may be built. To choose one of these
graphs as representing the actual network behaviour therefore requires
additional information. This additional information may simply be, for instance,
the datum specifying that all elementary network transitions are possible, or at
least that no elementary network transitions are known or considered to be
impossible. In this case, either the \GIG{} or the \GGIG{} (which can both be
built in time ${\cal O}(n\cdot 2^{2n})$) need to be chosen. They are the \TG{s}
that represent the alleged network behaviour the most completely. If the network
is known to perform only asynchronous transitions, then the \SIG{}, the \SSIG{}
(which can both be built in time ${\cal O}(n\cdot 2^{n})$) or sub-graphs of
these are better suited. And for networks that are supposed to be updated with a
given update schedule $\delta$ of period $p$, either $\T{\delta}^{\mbox{\tiny
    \sc elem}}$ or $\T{\delta}$ (which can both be
built in time ${\cal O}(np\cdot 2^{n})$), must be considered. Thus, in short, in
addition to the knowledge of the set of \LTF{s}, to derive the \TG{} of a
network, it is necessary (but not always sufficient) to specify the
\emph{nature} of the possible transitions or paths of the network. Without any
such indication, the network behaviour cannot be inferred non-ambiguously even
from an exhaustive knowledge of its underlying mechanisms (\ie, its structure
and interactions).\smallskip

\noindent Conversely, knowing the \TG{} of the network is not sufficient to
infer its set of \LTF{s} and its
structure~\cite{Robert1995,Robert1986,Tournier2005,Noual2011b}. In some cases,
however, with some simple additional information, it
is~\cite{Liang1998,Tournier2009}. Indeed, first, suppose that the network
behaviour is known to be fully described by a \emph{deterministic} \STS{}
$\T{}=(\Bn,T)$ in which all nodes have out-degree at most $1$ and all
transitions are of the form $(x, F(x))$ for a certain \GTF{}
$F:\Bn\to\Bn$. Then, from Equation~\ref{EQ-GTtofiPAR} below, the set of \LTF{s}
$\F=\{f_i\ |\ i\in V\}$ of the network can be derived in time ${\cal O}(n\cdot
2^n)$, \ie, in linear time with respect to the size of
$\T{}$~\cite{Tournier2005}.
\begin{equation}
  \label{EQ-GTtofiPAR}
  \forall i\in V,\ f_i: \left\{\begin{array}{rcl} 
      \Bn & \to & \B\\ 
      x & \mapsto & F(x)_i
    \end{array}\right.\text{.}
\end{equation}
If $\T{}$ is known to be a sub-graph of the \SIG{} or the \SSIG{}, then, also in
linear time with respect to the size of $\T{}$ (${\cal O}(n^2\cdot 2^n)$), $\F$
can be built using:
\begin{equation}
  \label{EQ-GTtofiSEQ}
  \forall i\in V,\ f_i: \left\{\begin{array}{rcl} 
      \Bn & \to & \B\\
      x &\mapsto & \begin{cases}
        \neg x_i & \text{if } (x,\overline{x}^{i})\in T\text{,}\\
        x_i & \text{otherwise\text{.}}
      \end{cases}
    \end{array}\right.
\end{equation}
More generally, under the hypothesis that \emph{all transitions in $\T{}$ are
  elementary}, the set $\F$ of \LTF{s} of the network can be derived in time
${\cal O}(n\cdot 2^{2n})$ by exploiting the following equation:
\begin{equation}
  \label{EQ-GTtofi}
  \forall i\in V,\ f_i: \left\{\begin{array}{rcl}
      \Bn & \to & \B\\
      x &\mapsto & \begin{cases}
        \neg x_i & \text{if } \exists y\in \Bn,\ (x,y)\in T \text{ and } 
        y_i\neq x_i\text{,}\\
        x_i & \text{otherwise.}
      \end{cases}
    \end{array}\right.
\end{equation}
In particular, Equation~\ref{EQ-GTtofi} can be used when the network is supposed
to be updated in parallel. For other \BS{} and \Sus{s}, Algorithm~\ref{ALGO-BS}
builds a set of \LTF{s} $\F$ from the dual input information of a \TG{} $\T{}$
and an \us{} $\delta$. To do so, it requires that $\T{}$ be indeed the graph of
a function $F:\Bn\to\Bn$ so that $\T{} = \T{\delta}$ and $F = F[\delta]$ be
possible. In addition, importantly, $\delta$ must be a \Sus{}. The reason for
this restriction is that Algorithm~\ref{ALGO-BS} (just as an observer would need
to do) relies on the knowledge that every apparent behaviour has no hidden
cause. If the transition $x\transRT y$ is observed, then, on the one hand, all
automata that have not \emph{apparently} changed states ($y_i = x_i$) have not
\emph{effectively} done so. If they had, they would have had to change an even
number of times in order to come back eventually into their initial state
($x_i$). This is not possible when automata are updated at most once between two
observations of the network configuration. On the other hand, if the automata
that do change states have not done so as a result of a series of unobservable
causes, then their changes of states can indeed be exploited to derive the
interactions which are their direct causes.
\begin{algorithm}[htbp!]
  \caption{From $\T{}$ and $\delta$ to $\F$ in time ${\cal O}(n^2\cdot 2^n)$}
  \label{ALGO-BS}
  \KwIn{
    \begin{itemize}
    \item A digraph $\T{}=(\Bn,T)$ in which each node has out-degree $1$ so that
      the only out-neighbour of any $x\in \Bn$ can be denoted by $F(x)$ and
    \item A \Sus{} $\delta\equiv (W_t)_{t\in \NN[p]}$.
    \end{itemize}
  } 
  \KwOut{
    A set of \LTF{s} $\F = \{f_i:\Bn\to \B\ |\ i\in V\}$
    such that $\T{} = \T{\delta}$ and $F = F[\delta]$.
  }\smallskip

  \noindent \ForAll{$x\in \Bn$}
  { $y\longleftarrow F(x)$\;
    \ForAll{$i\in W_0$}{ $f_i(x)\longleftarrow y_i$\; }
    \ForAll{$t<p$}
    { $x\longleftarrow F_{W_t}(x)$\;
      \ForAll{$i\in W_{t+1}$}{ $f_i(x)\longleftarrow y_i$\; }}}
\end{algorithm}

\noindent Let us highlight that \BSus{s} have the notable particularity of
allowing to observe the network configuration only once per period of updates
without loosing any crucial information about the nature of limit behaviours of
the network. It suffices to know the usually non-elementary\footnote{For \us{s}
  $\delta$ such as those that are considered in the present paragraph, unless
  the period of $\delta$ is $1$ as for the parallel \us{}, $\T{\delta}$ is
  non-elementary.} \TG{} $\T{\delta}$ to draw some significant information. For
example, consider a network updated with a \BSus{} $\delta\equiv (W_t)_{t\in
  \NN[p]}$ of period $p$. $\T{\delta}$ necessarily contains a sub-graph of
the following form, that is, a limit behaviour with a certain period
$k\in\N$:\\[2mm] \centerline{
  \begin{tikzpicture}
    \node[right] at (0.15,0) (x) 
    {$x\transRT F[\delta](x) \transRT F[\delta]^2(x)$~}; 
    \node[right] at (0,-1.3) (x1) 
    {$F[\delta]^{k-1}(x) \REVtransRT F[\delta]^{k-2}(x)\REVtransRT$} ;       
    \draw[-,dotted,thick,] (x) -- (8,0) -- (8,-1.34) -- (5.8,-1.34);
    \draw[->>, >=angle 60, right] (0.4,-0.95) -- (0.4,-0.25);
  \end{tikzpicture}
} Because the transitions involved in this behaviour are not necessarily
elementary, there might be more than $k$ events needed to loop on configuration
$x$. Precisely, there might be between $k$ and $p\times k$. Some of the
elementary updates required by $\delta$ along the closed path from $x$ to $x$
may be null so it does not hold that non-elementary limit behaviours of period
$k$ correspond to elementary limit behaviours of period $p\times k$.  However,
under \BSus{s}, it does hold that configurations $x$ that are observed to be
stable in $\T{\delta}$ (\ie, configurations that have null effective out-degree
and are fixed points of $F[\delta]$) really are stable: $\forall i\in V,\ i\in
\nU(x)$. In the general case, on the contrary, the \us{} might allow automata to
switch states several times between two network observations or it might omit
the update of an unstable automaton so as to give the impression that the
network is stable while it is not. Thus, unlike with other \us{s}, with
\BSus{s}, an observer that has knowledge of $\T{\delta}$ can distinguish stable
configurations from other limit behaviours.

\section{Theorisation and modelling of time}
\label{SEC-theorisation}

In this section we concentrate on the ``first step'' of the modelling process
which we call the \emph{theorisation} step. Its purpose is to define the
modelling framework for the modelling of a certain category of real systems such
as systems of genes that interact via their protein products.  In other words,
it aims at setting some grounds before any practical modelling of a real system
(this next step frames Section~\ref{SEC-effective-modelling}) can effectively be
done.  Theorisation thus first needs to choose a formal language in order to
describe the features of ``reality'' that are considered%
 \footnote{Let us note that the very act of describing mentally some
   observations is already in itself an automatic and often subconscious
   theorisation of reality so there obviously are some theorisations occuring at
   a ``lower-level'' than the level which is the object of this
   section. However, we ignore them here because their identification lies
   outside the scope of our competences (it requires, in particular, to have a
   good idea of what ``reality'' is precisely).}.  In our context, this language
 consists in the mathematical language that allows to express definitions
 relative to \BAN{s}. Next, or simultaneously, in this language, a \emph{theory}
 is defined. In consistency with one another, definitions are given to formal
 objects. Properties of these and relationships between them are specified. This
 leads, for instance, to the theory of \BAN{s} as described in
 Section~\ref{SEC-generaldefinitions}. From the language and theory that are
 chosen follows immediately a correspondence between features of reality
 intended to be modelled and features of the theory supposed to model them.
 This reality/theory correspondence is dually composed of a \emph{modelling
   map} \label{RTmaps} (which, informally represents the \emph{reality $\to$
   theory} direction of the correspondence) and an \emph{interpretation map}
 (representing the opposite direction, \emph{theory $\to$ reality}). The
 \emph{modelling map} specifies how portions of reality are represented
 mathematically. For example, it may specify that genes are modelled by
 automata, that interactions between genes via the proteins they code for are
 represented by Boolean functions, that changes of protein concentrations in the
 cell are simulated by transitions of a \TG{}\ldots Conversely, the
 \emph{interpretation map} associates a ``modelling meaning'' or a justification
 (possibly void) to each object and property of the theory. For example,
 according to this map, automata may be interpreted as genes, Boolean vectors
 may be interpreted as cell configurations and a local transition function $f_1:
 x\mapsto \neg x_0\wedge x_1$ can be regarded as modelling the fact that gene
 $G_1$ remains expressed only if gene $G_0$ is not, given that automaton
 $i\in\{0,1\}$ models gene $G_i$. Further, the ``interpretation map'' can also
 specify that a formal hypothesis stating that synchronous transitions are
 impossible (see Hypothesis~\ref{HYP-asynchrone} below) is the formal
 translation of the fortuitousness of two events ending
 simultaneously.\smallskip

\noindent The main difficulty of the theorisation step of modelling is the
definition of both the modelling and the interpretation maps. Indeed, from the
deliberately designed backbone of the reality/theory correspondence, it often
follows some complex and subtle ramifications. These ramifications can impose
implicitly that certain features of reality be matched to some precise features
of the theory and conversely. But they can also, on the contrary, forbid some
matches.  Further, generally, neither the modelling map nor the interpretation
map, are surjective: not all aspects of the theory can necessarily be
interpreted as the representation of something real and, obviously, not all
aspects of reality are represented by the theory.  As a consequence, the
modelling framework itself requires thorough coherence and bounding in its
definition.  \emph{Modelled features} of reality need to be identified and
distinguished from its \emph{non-modelled features}.  Dually, 
\emph{modelling parameters} and \emph{properties} of a theory, which can
reasonably be considered as representations of some portion of observed reality,
also need to be identified and distinguished from its \emph{non-modelling
  features} (those that are artefacts of formalisation rather than
pertinent representations of anything real).  The present section focuses on the
notion of time to highlight these difficulties and the coherence that is
required between the distinct associations specified or implied by the
reality/theory correspondence.\medskip

\noindent The very concept of \emph{transition} from one network state $x$ to another
$y$ suggests a notion of time that positions $x$ \emph{before} $y$.  The term
\emph{trajectory} that is usually used instead of \emph{network path} (see
Section~\ref{SEC-Transitions}) re-enforces the natural association that can be
made between an intuitive idea of time flow and a mathematical concept of causal
precedence. The \emph{length} of a series of transitions from $x$ to $y$
evokes the \emph{time} that the system spends to go from one point $x$ to
another $y$ in its state space.  And, especially in a modelling context where
the aim precisely is to relate experiences of reality and abstract concepts, the
formal language used tends to adapt to implicit associations that are made to
better understand theoretical objects and their properties.  The question of the
pertinence of these associations, however, is not always obvious. In the
sequel, we first recall some definitions on \DS{s} and show how \BAN{s} with a
given behaviour can be seen in terms of this formalism. Then, we discuss how, in
different formalisms, different points of view on the way time is taken into
account in a model give rise to different problems and questions in its
theoretical analysis. The section ends with an example that illustrates some of
these questions and in which \BAN{s} are seen as models of genetic regulation
networks.

\subsection{Dynamical systems}
\label{SEC-DS} 

A discrete-time dynamical system, called simply \emph{dynamical system} in the
sequel\footnote{We bypass the difficulty of choosing a time space $\t$ by
  assuming it is discrete. Continuous-time \DS{s} will thus not be mentioned at
  all. This choice is supported by the state space being discrete in our
  framework.} is a triplet $\D = (S, \t, \phi)$ where $S$ is the \emph{state
  space} of the system, $\t \subseteq \N$ is its \emph{time domain} (or
\emph{evolution space}) and $\phi : S \times \t \to S$ is the \emph{evolution
  function} describing the system dynamics. It satisfies:
\begin{equation*}
  \forall s\in S,\ \forall t_1,t_2\in \t,\ \phi(s,0) = s \text{ and }
  \phi(\phi(s,t_1),t_2) = \phi(s,t_1+t_2)\text{.}
\end{equation*} 
$\phi(s,t)$ represents the state of the system at time $t$ so that the
\emph{trajectory} (or path) of $\D$ initiated in state $s \in S$ is $\{\phi(s,
t)\ |\ t \in \t\}$.  Let us note that for any initial state $s \in S$, the
function $\phi_s : t \mapsto \phi(s,t)$ associates to every time step $t$, a
\emph{unique} image $\phi(s, t)$. In particular, it defines the unique successor
$\phi(s,1)$ of $s$. This allows  for two types of \DS{s}: deterministic
and stochastic.  In the case of stochastic \DS{s}, however, the terminology
introduced above needs to be adapted so that $S$ rather denotes \emph{a set of
  probability laws} on the state space of the system:
\begin{equation*}
  S \subseteq \{\mu \in [0,1]^{2^n}\ |\ \sum_{x\in\States} \mu_x = 1\}\text{.} 
\end{equation*}
In this case, let $x(t)$ be the random variable corresponding to the system
state at time $t$ and let $\mu(t)$ be the \PL{} of $x(t)$. Then, $\phi$ is such
that $\mu(t) = \phi(\mu(0), t)$.

\subsubsection*{Deterministic \DS{s}}

As an example of deterministic \DS{s}, let us consider a network updated with a
\PERus{} $\delta \equiv (W_k)_{k\in\, \NN[p]}$. As discussed in
Section~\ref{SEC-GT}, there are several ways to describe the network behaviour
in this case. First, focusing globally on \emph{periods} of $\delta$, one may
consider the \TG{} $\T{\delta}$ in which every configuration has out-degree
$1$. The system is then context-free and can be defined as a
deterministic \DS{} $\D = (\Bn, \N, \phi)$ where: 
\begin{equation*}
  \forall x\in\Bn,\ \forall t\in\N,\ \phi(x, t) = F[\delta]^t(x)\text{.}
\end{equation*}
A second way to describe the network behaviour under $\delta$ is to decompose
the transitions $x \transRT F[\delta](x)$ into series of elementary transitions
\begin{equation*}
  x \Trans[W_t] F_{W_t}(x),\ t \in \NN[p]\text{.}
\end{equation*}
 This yields the \TG{} $\Telem$ which is
not a \STS{} like $\T{\delta}$ but in which every node still has out-degree
$1$. $\Telem$ defines a context-dependent system (whether or not $W_0$ can be
updated in configuration $x$, for instance, depends on the previous elementary
transition performed by the system) that can be seen as a deterministic \DS{}
$(\Bn, \N, \phi)$ where:
\begin{equation*}
  \forall x\in \Bn,\ \forall t = k\cdot p +d \equiv d\mod{p},\ 
  \phi(x,t) = F_{W_d}\circ \ldots \circ F_{W_1} \circ F_{W_0} \circ 
  F[\delta]^k\text{.}
\end{equation*}
In this case, as the definition of $\phi$ shows, there is no time-independent
(global transition) function $\phi_1$ that associates a
unique successor $\phi_1(x) = \phi(x, 1)$ to every configuration $x$.

\subsubsection*{Stochastic \DS{s}}

When the \TG{} $\T{} = (\Bn, T)$ describing the behaviour of a network has nodes
of out-degree greater than $1$, there generally is no obvious way of defining it
as a \DS{}. However, with some additional indications or hypotheses,
probabilities may be assigned to each transition of $\T{}$. This way, \T{} can
be seen as the graph of a Markov chain on $\Bn$ and the adjacency matrix of \T{}
can be turned into a \emph{stochastic transition matrix}, often called the
Markovian matrix, \ie, a matrix $P$ of dimension $2^n \times 2^n$ satisfying:
\begin{equation*}
  \begin{tabular}{lll} 
    \textit{(i)} & $\forall x, y \in \Bn,$ & $P_{x,y} \in [0,1] \text{,}$\\[4pt] 
    \textit{(ii)} & $\forall x \in \Bn,$ & $\sum_{y\in \Bn} P_{x,y} = 
    1 \text{ and }$\\[4pt] 
    \textit{(iii)} & $\forall x, y \in \Bn,$ & $(x,y) \notin T \Rar P_{x,y} = 
    0\text{.}$
  \end{tabular}
\end{equation*}
The component $P_{x,y}$ in this matrix represents the probability that the
network performs transition $(x, y)$, \ie, the probability that it reaches
configuration $y$ given that it was in configuration $x$ at the previous time
step: 
\begin{equation*}
  P_{x,y} = \proba \big(x(t+1) = y\ |\ x(t) = x\big)\text{.}
\end{equation*}
The probability that it changes configurations is given by: 
\begin{equation*}
  \proba \big( x(t+1) \neq x(t) \big) = \sum_{y\neq x\in\Bn} P_{x,y}\text{.}
\end{equation*}
Then, the network behaviour can be defined as a stochastic \DS{} whose evolution
function is given by:
\begin{equation*}
  \begin{cases}
    \phi(\mu,0)=\mu\text{,}\\
    \phi(\mu,t)=\mu\cdot P^t\text{,}
  \end{cases}
\end{equation*}
where $\mu$ is an arbitrary \PL{} on the state space of the system ($\mu \in
[0,1]^{2^n}$ and $\sum_{x\in\States} \mu_x = 1$).  Thus,  if $\mu = \mu(0)$
is the \PL{} of the initial network configuration $x(0) \in \Bn$, then $\mu(t) =
\phi(\mu,t)$ is the \PL{} of the network configuration $x(t) \in \Bn$ at time
step $t$ ($\forall x \in \Bn,\ \mu_x(t) = \proba \big( x(t) = x \big)$).
\def\gig{\text{\tiny \GIG}}
\begin{example}
  Consider the network of Examples~\ref{EX-structure-E1} to~\ref{EX-updates-E1}
  and its \GGIG{}  $\eff{\T{\smallg}} = (\B^3, T)$ and given in
  Example~\ref{EX-TG-E1}. Then, introducing a rate $\alpha \in [0,1]$ at which
  automata of the network are updated (at each time step, any automaton $i \in
  V$ is updated with probability
  $\alpha$)~\cite{Fates2006,Regnault2007,Rouquier2009}, the stochastic
  transition matrix $P$ can defined as follows:
  \begin{equation*}
    \forall (x,y)\in \B^3\times\B^3, P_{x,y} = \begin{cases}
      \alpha^{d_{xy}}\cdot (1-\alpha)^{\scriptstyle 
        {\cal U}\displaystyle_x- d_{xy}} & \text{ if } (x,y)\in T,\\ 
      0 & \text{otherwise,}
    \end{cases}
  \end{equation*} 
  where $d_{xy} = |D(x,y)|$ and $\scriptstyle {\cal U}\displaystyle_x=|\,
  \U(x)|$ (see Section~\ref{SEC-stability} for notations). This yields the
  following \TG{} where transitions of $\eff{\T{\smallg}}$ are labelled by their
  probabilities:\\[2mm]
  \centerline{
    \begin{minipage}{0.02\textwidth}
      ~~~
    \end{minipage}
    \begin{minipage}{0.95\textwidth}
      \centerline{
        \hspace*{18pt}\begin{tikzpicture}[descr/.style={fill=white,inner
        sep=2.5pt}][descr/.style={fill=white,inner sep=2.5pt}]
      \node at (0,0) (x0) {$(0,0,0)$~};
      \node at (0,2) (x1) {$(0,0,1)$~};
      \node at (9,0) (x2) {$(0,1,0)$~}; 
      \node at (9,2) (x3) {$(0,1,1)$~}; 
      \node at (3,0) (x4) {$(1,0,0)$~}; 
      \node at (3,2) (x5) {$(1,0,1)$~}; 
      \node at (6,0) (x6) {$(1,1,0)$~}; 
      \node at (6,2) (x7) {$(1,1,1)$~}; 
      \path[->,font=\scriptsize, >=angle 60,] (x0) edge[->] node[descr] 
      {$\alpha\overline{\alpha}$} (x1) edge[->] node[descr] {$\alpha^2$} 
      (x5) edge[->] node[descr] {$\alpha\overline{\alpha}$} (x4);
      \path[->,font=\scriptsize, >=angle 60,] (x1) edge[->] node[descr] 
      {$\alpha$} (x5);
      \path[->,font=\scriptsize, >=angle 60,] (x4) edge[->] node[descr]  
      {$\alpha\overline{\alpha}$} (x5) edge[->] node[descr] {$\alpha^2$} 
      (x7) edge[->] node[descr] {$\alpha\overline{\alpha}$} (x6);
      \path[->,font=\scriptsize, >=angle 60,] (x7) edge[->] node[descr] 
      {$\alpha$} (x6);
      \path[->,font=\scriptsize, >=angle 60,] (x2) edge[->] node[descr] 
      {$\alpha$} (x6);
      \path[->,font=\scriptsize, >=angle 60,] (x3) edge[->] node[descr] 
      {$\alpha\overline{\alpha}$} (x7) edge[->] node[descr] {$\alpha^2$} 
      (x6) edge[->] node[descr] {$\alpha\overline{\alpha}$} (x2);
      \draw (x0) to [->,font=\scriptsize, min distance=10mm, in =-135, 
      out=-45] node[below] {$\overline{\alpha}^2$} (x0);
      \draw (x1) to [->,font=\scriptsize, min distance=10mm, in =45, 
      out=135] node[above] {$\overline{\alpha}$} (x1);
      \draw (x2) to [->,font=\scriptsize, min distance=10mm, in =-135, 
      out=-45] node[below] {$\overline{\alpha}$} (x2);
      \draw (x3) to [->,font=\scriptsize, min distance=10mm, in =45, 
      out=135] node[above] {$\overline{\alpha}^2$} (x3);
      \draw (x4) to [->,font=\scriptsize, min distance=10mm, in =-135, 
      out=-45] node[below] {$\overline{\alpha}^2$} (x4);
      \draw (x5) to [->,font=\scriptsize, min distance=10mm, in =45, 
      out=135] node[above] {$1$} (x5);
      \draw (x6) to [->,font=\scriptsize, min distance=10mm, in =-135, 
      out=-45] node[below] {$1$} (x6);
      \draw (x7) to [->,font=\scriptsize, min distance=10mm, in =45, 
      out=135] node[above] {$\overline{\alpha}$} (x7);
    \end{tikzpicture}
      }
    \end{minipage}
    \begin{minipage}{0.02\textwidth}
      \vspace*{4.07cm}\eex
    \end{minipage}
  }
\end{example}

\subsection{Modelling time}
\label{SEC-time} 

In the case of networks seen as \DS{s} (deterministic or stochastic), because
the set $\t$ is called the time domain of the system and $t\in\t$ is called a
time step, a \emph{moment} or a \emph{date}, an implicit notion of time is
introduced that can be interpreted several ways.  In this section we discuss
three different points of view that can be taken, in the context of modelling,
on the abstractions of time arising from the definition of a network
behaviour. The first two points of view derive from the formalisms of
\DS{s}. The last one follows from that of \STS{s} which are called
``\emph{causal systems}'' here to emphasise their difference with
\DS{s}\footnote{Let us note that for similar reasons, that is, to emphasise the
  difference between the two main angles (causal and dynamical) that can be
  adopted to study \BAN{s}, in Section~\ref{SEC-generaldefinitions}, we have
  deliberately chosen to use the term \emph{behaviour} of a network rather than
  the terms \emph{dynamical behaviour} and \emph{dynamics}.}.

\subsubsection*{Modelling durations} 

A first reading of the mathematical concept of time defined by \DS{s} consists
in interpreting it as a literal match of the real time so that time steps in
$\t$ are taken as a unit of measurement of real time and all possible network
transitions are supposed to take the same amount of time, that is, one
unit. When the network behaviour is described by an elementary \TG{} \T{}, for
instance, all elementary updates are supposed to take the same time, whatever
the automata that they update.\smallskip

\noindent When the system movements modelled by transitions can however not all
be assumed to take the same fixed amount of time, $\t$ cannot be interpreted as
a discretised version of a real time flow. In that case, to maintain a modelling
of transition durations, the time domain must be backed up with some additional
parameters. One straightforward method is to label each transition of the
network by a value that measures the time taken by this transition, or rather,
by the event modelled by this transition. In these lines,
in~\cite{Thomas1973,Bernot2004,Thomas1983,Thomas1990,Kaufman1985,Siebert2006},
the authors have refined the formalism of \BAN{s} in order to model genetic
regulation networks and take into account some of the time delays to which
regulations are submitted (see Section~\ref{SEC-delays}).\smallskip

\noindent In addition to the questions that are mentioned in the next paragraphs
which are also natural and pertinent with more general approaches, this point of
view (like any point of view) on time yields a set of theoretical questions that
are specific to it.  These questions rely on the strong hypothesis that the
concept of time in the definition of a \DS{} is indeed meaningful in terms of
modelling. As an example, let us cite the following non-exhaustive list:
\emph{How long does the network take or is the network expected to take to reach
  a certain configuration or to start displaying a certain behaviour?}
\emph{What is the (most likely) network configuration that is reached in time
  $t$?}, \emph{When or how long will the network display this behaviour?},
\emph{How many times is the network expected to reach this configuration during
  this lapse of time?}\ldots

\subsubsection*{Modelling precedence}
\label{SEC-precedence}

A second reading of the mathematical concept of time inherent to \DS{s} consists
in understanding it as a simple \emph{evolution parameter} defining no more than
a relation of precedence between network configurations and without implying any
notion of duration. If the two transitions $x\transRT y$ and $x'\transRT y'$ are
both possible, then, with this point of view, it becomes coherent to accept that
$x\transRT y$ may take much longer to happen than $x'\transRT y'$, under certain
circumstances, while, perhaps, under different circumstances, the opposite is
true (\ie, $x'\transRT y'$ takes longer than $x\transRT y$). Thus, different
behaviours of the network can take place at different time scales although no
additional precisions aim at distinguishing these time scales or the different
possibilities they yield. The mathematical concept of time of \DS{s} is regarded
as a \emph{logical} version of time and it requires less information on the
nature of transitions and on how they happen. Paths or trajectories simply are
sequences of successive events.  The time they take cannot be measured but the
number of events or elementary events they involve can however be
counted. Consequently, the questions that characterise this point of view on
\BAN{s} and \DS{s} are of the following form: \emph{How many steps does the
  network take or is the network expected to take to reach a certain
  configuration or to start displaying a certain behaviour?} \emph{What is (the
  most likely) network configuration that is reached after $k$ steps?},
\emph{Can a given behaviour be observed \emph{after} a certain other?}, \emph{What
  trajectories or behaviours are more likely?}\ldots

\subsubsection*{Modelling causes}
\label{SEC-causal}

One last point of view consists in ignoring altogether any associations that can
be made between an intuitive idea of time (precedence as well as duration) and
the theoretical features that follow from formalisation.  This way, contrary to
the case where \BAN{s} are assimilated to \DS{s}, no more information than the
\TG{} is required. Any arbitrary \BAN{} can be regarded as a \emph{causal
  system} (\ie, a \STS{}) stripped from any notion of time. Obviously, there is
no notion of duration associated to causal systems. But neither is there any
meaningful notion of time precedence that follows naturally and non ambiguously
from their definition. Indeed, when several transitions $(x, y^k),\ k\in \N$,
are possible in the same network configuration $x$, then none of the network
configurations $y^k$ is the configuration that is reached after $x$. Each
configuration $y^k$ is only the result of one of several possible events which
may occur with an unknown probability and within an unknown lapse of time. The
notion of \emph{moment} is therefore replaced by the notion of
\emph{possibility} and duration is replaced by a logical relation between causes
and consequences. Two transitions $(x,y)$ and $(x,z)$ being possible means that
$x$ allows at least two different ``\emph{continuations}'', $y$ and $z$. Causal
systems are non-temporised systems in which the focus is placed on the degrees
of freedom (\ie, the set of possible state switches) that the network and the
automata in it have in each configuration.  Time-related questions such as those
that have been mentioned in the previous paragraphs loose their immediate
meaning. Further, although the problem of how to prune a \TG{} in order to make
all trajectories deterministic can obviously be pertinent in the case of a
dynamical system (provided additional information, such as a stochastic
transition matrix $P$ that specifies what transitions can indeed be ruled out),
it is not in the more general context of \STS{s}. Indeed, in this context,
transitions are associated to no more information than that of there own
existence. Thus, no transition of the system that is \emph{a priori} possible
can be disregarded \emph{a posteriori}, even if it represents a highly
improbable event. The only pertinent questions in the context of \STS{s} are
``existence questions'' such as: \emph{Can a configuration that satisfies a
  certain set of properties be reached from a given configuration $x\in\Bn$?},
\emph{Is a given behaviour possible?}, \emph{Can this transition be made?}, {\em
  What new behaviours can be reached or become possible if some new transitions
  are added?}\ldots One example of a problem that fits exactly into this
framework is studied in~\cite{Noual2011b}. The main question it addresses is
whether the existence of synchronous transitions in a \TG{} increases or
decreases the possibilities in the network behaviour.

\subsection{Example: modelling time and genetic regulation networks}
\label{SEC-delays}

In~\cite{Thomas1995,Thomas1983,Kaufman1985}, the authors develop a formalism
based on \BAN{s} to model genetic regulation networks. This formalism integrates
a formal notion of delay to account for the time flow that the regulations are
subjected to in reality. It also assumes the two following statements:
\begin{hyp} 
  Only asynchronous transitions are possible.
  \label{HYP-asynchrone}
\end{hyp}
\begin{hyp} 
  All effective asynchronous transitions that update and activate
  (resp. deactivate) the same automaton take the same amount of time.
  \label{HYP-same-transition-time}
\end{hyp}
More precisely, each automaton $i \in V$ is assigned two values $\delup \in
\mathbb{R}$ and $\deldwn \in \mathbb{R}$, called respectively its
\emph{activation} and \emph{deactivation delays}. They represent the time it
takes for automaton $i$ to be updated and to change states: $\delup$
(resp. $\deldwn$) corresponds to the time $i$ takes to switch from $0$ to $1$
(resp. from $1$ to $0$). By Hypothesis~\ref{HYP-same-transition-time}, 
delays do not depend on the context: whatever the network configuration $x \in
\Bn$ in which $i$ is unstable ($i \in \U(x)$), transition $x
\Seq[i]\overline{x}^{i}$ lasts $t_i$ time units where $t_i = \delup$ if $x_i =
0$ and $t_i = \deldwn$ if $x_i = 1$. By Hypothesis~\ref{HYP-asynchrone}, this
covers all possible transitions so, in the rest of this section, we use the
following notations:
\begin{equation*}
  x\Seq[\delup]\overline{x}^{i} \text{ if } x_i = 0 \quad\text{and}\quad 
  x\Seq[\deldwn]\overline{x}^{i} \text{ if } x_i = 1\text{.}
\end{equation*}
\begin{figure}[t!]
  \centerline{\scalebox{0.71}{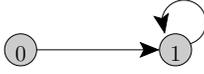}}
  \caption{Interaction graph of the network considered in
    Example~\ref{EX-contrex}.}
  \label{FIG-struct-contrex}
\end{figure}
\begin{example}
  \label{EX-contrex}
  Consider a network of size $2$ whose structure is pictured by
  Figure~\ref{FIG-struct-contrex} and whose set of \LTF{s} is:
  \begin{equation*}
    f_0:\fcn{\B^2}{\B}{x}{1} \quad \text{ and } \quad 
    f_1:\fcn{\B^2}{\B}{x}{\neg x_0 \vee x_1}\text{.}
  \end{equation*}
  The \SSIG{} of this network, completed with the delay specifications, is
  given below:\\[2mm]
  \centerline{
    \begin{minipage}{0.02\textwidth}
      ~~~
    \end{minipage}
    \begin{minipage}{0.95\textwidth}
      \centerline{
        \begin{tikzpicture}[descr/.style={fill=white,inner sep=2.5pt}]
          \node at (0,0)   (zero)  {$(0,0)$}; 
          \node at (2.5,0) (un)    {$(0,1)$}; 
          \node at (0,2)   (deux)  {$(1,0)$};
          \node at (2.5,2) (trois) {$(1,1)$};
          \path (zero) edge[-open triangle 60] node[descr] 
          {$\scriptstyle \delup[0]$} (deux);
          \path (zero) edge[-open triangle 60] node[descr] 
          {$\scriptstyle \delup[1]$} (un);
          \path (un) edge[-open triangle 60] node[descr] 
          {$\scriptstyle \delup[0]$} (trois);
          \draw (deux) to [->, >=angle 60, min distance=10mm, in =90, out=180] 
          node[descr] {$\scriptstyle 0,1$} (deux);
          \draw (trois) to [->, >=angle 60, min distance=10mm, in =90, out=0] 
          node[descr] {$\scriptstyle 0,1$} (trois);
          \draw (un) to [->, >=angle 60, min distance=10mm, in =-90, out=0] 
          node[descr] {$\scriptstyle 1$} (un);
        \end{tikzpicture}
      }
    \end{minipage}
    \begin{minipage}{0.02\textwidth}
      \vspace*{3.2cm}\eex
    \end{minipage}
  }
\end{example}

\noindent An appreciable consequence of the addition of delays, in a modelling
context, and one of their main purposes is to allow a natural deterministic
interpretation of a network behaviour. Indeed, when two transitions are
possible in a same configuration, we may suppose that the one that does
effectively take place is the one that happens faster. Thus, in configuration
$(0,0)$ of the network of Example~\ref{EX-contrex}, for instance, if $\delup[0]
< \delup[1]$, then transition $(0, 0) \Seq[{\delup[0]}] (0,1)$ can be supposed
to be the one to occur. Otherwise, if $\delup[0] > \delup[1]$, transition $(0,0)
\Seq[{\delup[1]}] (0,1)$ can be considered as the most likely.

\subsubsection*{Interpreting asynchronicity}

Let us analyse the meaning of Hypothesis~\ref{HYP-asynchrone} with respect to
the formalism of \BAN{s}. To do so, let $x\in\Bn$ be a network configuration in
which several automata are unstable. In particular, let $i,j\in \U(x)$.  By
definition, both automata $i$ and $j$ are ``on the verge of'' changing states in $x$
but by Hypothesis~\ref{HYP-asynchrone}, they cannot \emph{both} change states in
that configuration. In other terms, transitions
\begin{equation*}
  x\seq[i] \overline{x}^{i} \quad\text{and}\quad x \seq[j] \overline{x}^{j}
\end{equation*} 
are both possible but transition
\begin{equation*}
  x\pll[\{i,j\}] \overline{x}^{\{i,j\}}
\end{equation*} 
is not. This apparent conflict between the predefined theory of \BAN{s} (which
specifies that both automata can change states and have no reason not to) and
Hypothesis~\ref{HYP-asynchrone} (which disallows them from changing states under
some circumstances) can be addressed by additional notions of time and
duration. The delays implied by Hypothesis~\ref{HYP-same-transition-time} and
introduced in~\cite{Thomas1995,Thomas1983,Kaufman1985} provide these notions
precisely. Indeed, labelling transitions by delays augments their meaning beyond
that of a simple relation of precedence or causality as discussed in
Section~\ref{SEC-time}. And with this additional meaning, rather than stating
that $i$ and $j$ cannot both change states, Hypothesis~\ref{HYP-asynchrone} can
be understood as imposing that $i$ and $j$ cannot both change states
\emph{simultaneously}. It becomes significant and coherent to consider what
happens \emph{during} a transition $x\seq y$, between the moment where the
network is in configuration $x$ and the moment it reaches configuration
$y$. This way, more precisely still, Hypothesis~\ref{HYP-asynchrone} can be
taken as the claim that no two automata can \emph{finish} changing states
simultaneously.\smallskip

\noindent With this interpretation of Hypothesis~\ref{HYP-asynchrone}, let us
suppose that in configuration $x$, $i$ and $j$ are respectively subjected to the
delays $\del_i \in \{\deldwn, \delup\}$ and $\del_j \in \{\deldwn[j],
\delup[j]\}$ and that:
\begin{equation*}
  \del_i < \del_j\text{.}
\end{equation*} 
In configuration $x$, say at time $0$, both automata start changing states
since, according to the theory, they can. However at time $\del_i$, automaton
$i$ has effectively changed states whereas automaton $j$ has not yet had the
time to. At this moment, only two situations are \emph{a priori} coherent with
both the theory of \BAN{s} and our interpretation of
Hypothesis~\ref{HYP-asynchrone}:
\begin{enumerate}
\item Either $j$ has become stable ($j \in \nU(\overline{x}^{i})$) in which case
  it must be that $j$ is influenced (directly or indirectly) by $i$ (\ie, there
  is an arc or a path from $i$ to $j$ in the network interaction structure);
\item Or $j$ is not influenced (neither indirectly nor directly) by $i$ (and nor
  is its instability) and thus, the change of states of $i$ (effective at the
  current time $\del_i$) has not affected $j$ which can still change states in
  $\overline{x}^{i}$ ($j\in \U(\overline{x}^{i})$).
\end{enumerate}
The second case requires more attention and supplementary
justifications. Indeed, in this case, Hypothesis~\ref{HYP-asynchrone} imposes
that the trajectory:\\[2mm]
\centerline{
  \begin{tikzpicture}[descr/.style={fill=white,innersep=2.5pt}] 
    \node at (0,0)   (x) {$x$}; 
    \node at (2.5,0) (y) {$\overline{x}^{i}$};
    \node at (5,0)   (z) {$\overline{x}^{\{i,j\}}$}; 
    \path (x) edge[-open triangle 60] node[above] {$\scriptstyle \del_i$} (y); 
    \path (y) edge[-open triangle 60] node[above] {$\scriptstyle \del_j$} (z);
  \end{tikzpicture}
} is possible while the trajectory:\\[2mm]
\centerline{
  \begin{tikzpicture} 
    \node at (0,0) (x) {$x$}; 
    \node at (3,0) (y) {$\overline{x}^{i}$}; 
    \node at (6,0) (z) {$\overline{x}^{\{i,j\}}$}; 
    \path (x) edge[-open triangle 60] node[above] {$\scriptstyle \del_i$} (y); 
    \path (x) edge[-open triangle 60, out=-30,in=-155] node[above] 
    {$\scriptstyle \del_j$} (z);
  \end{tikzpicture}
} is not. This means that even though $j$ starts changing states in $x$, it must
stop and start all over again in $\overline{x}^i$, making the whole process last
$\del_i + \del_j$ time units rather than $max\{\del_i,\del_j\} =
\del_j$. Consequently, the change underwent by $i$ in transition $x \Seq[i]
\overline{x}^{i}$ which \emph{a priori} involves $i$ alone has an impact on the
possibilities of $j$. And this impact is non-negligible since it concerns the
duration of an event in a context where durations, precisely, have been given
significance.  There are two ways to preserve the consistency of the fact that
$i$ thus influences $j$ in all cases and the formal basis of the theory of
\BAN{s}. One can either suppose that two automata cannot be simultaneously
unstable unless they have a mutual and symmetric influence on one another:
\begin{equation*}
  \forall i,j\in V,\ \exists x\in\Bn,\ \{i,j\}\in \U(x)\ \implies\
  (i,j)\in A \text{ and } (j,i)\in A\text{.}
\end{equation*}
Or one can rule out simultaneity altogether and take on the following
hypothesis:
\begin{hyp}
  No two distinct events can start, finish or occur synchronously.
  \label{HYP-no-simult}
\end{hyp}
Because the first solution is very restrictive, we choose the second and for the
rest of this section, we accept Hypothesis~\ref{HYP-no-simult}. Let us, however,
point out that this snag introduced by Hypothesis~\ref{HYP-asynchrone}
highlights the difficulty, in a modelling context, of justifying restrictions or
refinements brought to a theory. In particular, it shows that choosing to
exclude some elementary transitions in the description of the behaviour of a
\BAN{} is a non-trivial choice when the aim is not simply theoretical
convenience.  Indeed, to defend this choice some strong arguments are needed
(such as those required to justify Hypothesis~\ref{HYP-no-simult} which is
needed to support Hypothesis~\ref{HYP-asynchrone}, as we have seen above) that
can obviously not be drawn from the theory itself but must still remain coherent
with it.

\subsubsection*{Genetic regulation systems and \BAN{s} with delays}

To study how \BAN{s} with delays model genetic regulation systems, let us now
consider a simple genetic regulation system involving exactly $n$ distinct genes
$G_0, \ldots, G_{n-1}$. Very schematically, each gene $G_i$ may undergo a
process by which the information it carries is ``decoded'' and
``read''~\cite{Jacob1961,Crick1970} to induce the synthesis of one or several
proteins. For each gene $G_i$, let $P_i$ be one of these proteins products of
$G_i$\footnote{We consider only one protein for each gene for the sake of
  simplicity. Considering several would be more realistic and would not alter
  the basis and thrust of our argumentation. But it would considerably burden
  its formulation.}. When $G_i$ is being decoded and read, we say that it is
\emph{active}. The activation of genes $G_i, \ldots, G_{n-1}$ can be modelled by
$n$ Boolean variables $g_0, \ldots, g_{n-1}$ so that, $\forall i \in V = \{0,
\ldots, n-1\}$, $g_i = 1$ (resp. $g_i = 0$) represents the activation (resp. the
non-activation) of gene $G_i$.\smallskip

\noindent Proteins can regulate the activation of genes. Thus, a gene $G_i$, via
its protein $P_i$, may have a retro-action on the system of genes to which it
belongs. For instance, it may influence the activation of gene $G_j$. If it
does, this (indirect) regulation of $G_j$ by $G_i$ can be represented by an arc
$(i, j) \in A$ of an interaction graph $\G = (V,A)$. More precisely, and with
respect to other protein influences that can act on $G_j$, it can be represented
by a local transition function $f_i : \Bn \to \B$.\smallskip

\noindent The effect of a regulation modelled by an arc $(i, j) \in A$ is
however subjected to there being enough proteins $P_i$ in the cell.  Here, for
the sake of simplicity, we suppose that each protein $P_i$ becomes able to
regulate the activation of any other gene if and only if its concentration
exceeds a certain threshold. If it does, the protein is said to be
\emph{active}. This hypothesis precisely allows a Boolean modelling of genetic
systems: a protein can either be active with respect to all influences it may
have on the system, or it may be inactive for all of them (if several levels of
activation of $P_i$ needed to be considered, it could not be modelled by just
two states).\smallskip

\noindent 
Let us say that gene $G_i$ is \emph{expressed} when protein $P_i$ is active. And
let us highlight the difference in our use of the two terms \emph{active} and
\emph{expressed} concerning genes. Gene $G_i$ may be active without being
expressed. This usually happens right after the precise instant $G_i$ is
activated. Before it becomes expressed as well, that is, before the precise
instant the concentration in $P_i$ hits its activation threshold, a time lapse
is required during which the concentration in $P_i$ increases but remains under
its activation threshold. This time lapse, precisely, is modelled by the delay
$\delup$. Conversely, $G_i$ may be inactive and expressed. Indeed, following the
instant $G_i$ is deactivated, it may remain expressed during some time. This
time corresponds to the time it takes for the concentration in $P_i$ to decrease
as a result of the molecules degradation and of their not being renewed (since
the synthesis of $P_i$ is no longer ``commanded'' by $G_i$), and fall below its
activation threshold. It is modelled by the delay $\deldwn$.  In summary, delays
$\delup$ and $\deldwn$ both represent the time that $P_i$ takes to respond
accordingly to a ``command'' sent by $G_i$.\smallskip




\noindent More formally, let us introduce $n$ new Boolean variables, $x_0,
\ldots x_{n-1}$, to model the expression of each gene $G_i$, or equivalently, to
model the activation of each protein $P_i$: $\forall i \in V,\ x_i = 1$
(resp. $x_i = 0$) models the expression of $G_i$ (resp. its non-expression) and
the activation of protein $P_i$ (resp. its non-activation).  Then, to model the
regulation system comprised of genes $G_i$ and their products $P_i$, let us
consider a \BAN{} of size $n$, whose interaction structure and \LTF{s} are as
suggested above and whose automata states are the variables $x_i,\ i \in
V$. Thus, a network automaton $i\in V$ rather represents the protein $P_i$ than
its coding gene $G_i$\footnote{This is only a convention that we set and which
  follows naturally from our choices of notation. Other presentations that lead
  to automata corresponding to genes rather than to proteins are possible.}. Let
us consider a situation in which the activations of genes $G_i$ are modelled by
the vector $g = (g_0, \ldots, g_{n-1}) \in \Bn$ and the activations of the
proteins $P_i$ are represented by the network configuration $x = (x_0, \ldots,
x_{n-1}) \in \Bn$. To model such a situation, we use the following matrix which,
abusing language, we call the \emph{network configuration}:
\begin{equation}
  \label{EQ-newconfig}
  \PG{x}{g}= \left[\begin{array}{@{\hspace{4pt}}c@{~\ldots~}c@{\hspace{3pt}}}
      x_0 &  x_{n-1}\\
      g_0 &  g_{n-1}
    \end{array}\right]\text{.}
\end{equation}

\noindent In~\cite{Thomas1995,Thomas1983,Kaufman1985}, only configurations where $g_i = f_i(x)$ are
considered, \ie, Hypothesis~\ref{HYP-fortuitintheface} figuring below is
made. As a consequence, $g_i = x_i$ is equivalent to automaton $i$ being stable
($i \in \nU(x)$) and, $\forall i \in V,\ g_i = x_i$ is equivalent to the
configuration being a stable configuration.
\begin{hyp}
  The only possible network configurations have the following form:
  \begin{equation*}
    \left[\begin{array}{@{\hspace{4pt}}c@{~\ldots~}c@{\hspace{3pt}}}
        x_0 & x_{n-1}\\
        g_0 & g_{n-1}
      \end{array}\right] = 
    \left[\begin{array}{@{\hspace{4pt}}l@{~\ldots~}l@{\hspace{3pt}}}
        x_0 &  x_{n-1}\\
        f_0(x) &  f_{n-1}(x)
      \end{array}\right]\text{.}
  \end{equation*}
  \label{HYP-fortuitintheface}
\end{hyp}

\begin{example}
  \label{EX-contrex-newconfig}
  With the notation introduced in Equation~\ref{EQ-newconfig} and with
  Hypothesis~\ref{HYP-fortuitintheface}, the \TG{} of Example~\ref{EX-contrex}
  still has only four configurations, those that satisfy $g_0 = 1$ and $g_1 =
  \neg x_0 \vee x_1$:\\[2mm]
  \centerline{
    \begin{tikzpicture}[descr/.style={fill=white,inner sep=2.5pt}]
      \node at (0,0)     (zero)  {$\PGPG{0}{1}{0}{1}$}; 
      \node at (3.1,0)   (un)    {$\PGPG{0}{1}{1}{1}$};           
      \node at (0,2.2)   (deux)  {$\PGPG{1}{1}{0}{0}$}; 
      \node at (3.1,2.2) (trois) {$\PGPG{1}{1}{1}{1}$}; 
      \path (zero) edge[-open triangle 60] node[descr] 
      {$\scriptstyle \delup[0]$} (deux);
      \path (zero) edge[-open triangle 60] node[descr] 
      {$\scriptstyle \delup[1]$} (un);
      \path (un) edge[-open triangle 60] node[descr] 
      {$\scriptstyle \delup[0]$} (trois);
      \draw (deux) to [->,>=angle 60, min distance=10mm, in =90, out=180]
      node[descr] {$\scriptstyle 0,1$} (deux);
      \draw (trois) to [->, >=angle 60, min distance=10mm, in =90, out=0] 
      node[descr] {$\scriptstyle 0,1$} (trois);
      \draw (un) to [->, >=angle 60, min distance=10mm, in =-90, out=0] 
      node[descr] {$\scriptstyle 1$} (un);
    \end{tikzpicture}
  } In particular, the following network configurations are supposed to be
  unrealisable:\\[2mm]
  \begin{minipage}{0.02\textwidth}
    ~~~
  \end{minipage}
  \begin{minipage}{0.95\textwidth}
    \begin{equation*}
      \PGPG{1}{1}{1}{0} \quad\text{and}\quad \PGPG{1}{1}{0}{1}\text{.}
    \end{equation*}
  \end{minipage}
  \begin{minipage}{0.02\textwidth}
    \eex
  \end{minipage}
\end{example}

\begin{figure}[h!]
  \centerline{\scalebox{0.7}{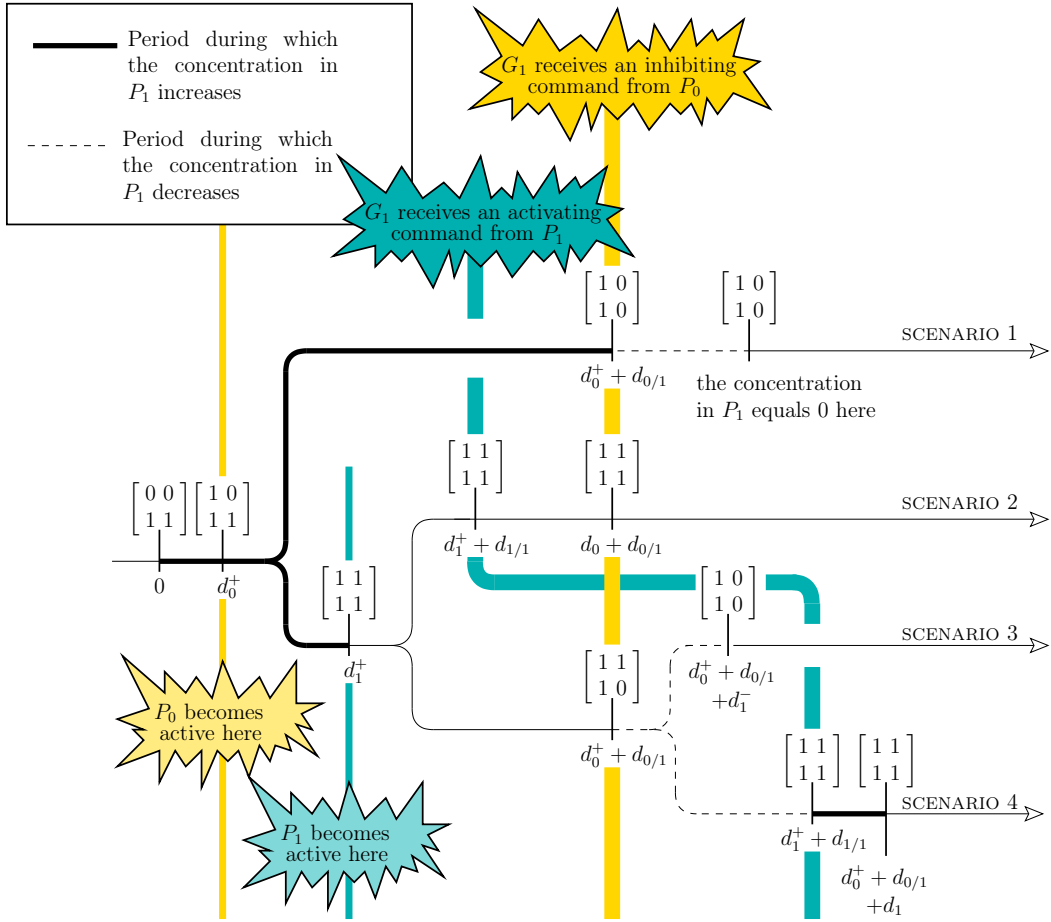}} 
  \caption{Diagram representing possible behaviours of the regulation network of
    Examples \ref{EX-contrex} and \ref{EX-contrex-newconfig} under the
    assumption that $\delup[0] < \deldwn[1]$. Let us emphasise that no effort
    whatsoever has been put into representing the time scale. Consequently,
    despite the space that separates the representation of the dates $\delup[0]$
    and $\delup[0] + \delPG{0}$, it is not excluded that $\delup[0] \approx
    \delup[0] + \delPG{0}$.}
  \label{FIG-contrex}
\end{figure}
\noindent \BAN{s} with delays can thus serve as models in which a mathematical
notion of time flow and duration is introduced to match the time that rules over
the interactions of the real system they represent. As mentioned in the previous
paragraphs, it is important to note, however, that the pertinence of the
modelling strongly depends on the hypotheses that are added to the original
theory. More specifically, it depends on the interpretations of these hypotheses
and on their consistency with respect to prior interpretations of other features
of the theory. We have already discussed above Hypothesis~\ref{HYP-asynchrone}
and how its significance relies on the additional notion of time introduced by
means of delays, and more precisely, how it relies on
Hypothesis~\ref{HYP-no-simult} (unless, as we have demonstrated, all network
automata are supposed to interact two by two). We propose now to analyse further
the role of hypotheses and the correspondence between this refined version of
the theory of \BAN{s} and the genetic regulation systems that they are intended
to model.

\subsubsection*{A rigorous adaptation and interpretation of \BAN{s} with delays}


By Hypothesis~\ref{HYP-no-simult} (and {\it a fortiori} by
Hypothesis~\ref{HYP-asynchrone}), we have assumed that proteins do not receive
commands from their coding genes \emph{immediately} when these are emitted. For
the precise same reason (\ie, two distinct punctual events are necessarily
separated in time by a lapse of time, however small it be), let us suppose
similarly that genes cannot change states \emph{simultaneously} to the changes
of the states of their regulating proteins.  Of course, one can argue that the
two phenomena evoked are set in different time scales and that the second
direction involves times lapses that are completely insignificant compared to
those involved by the first. Nevertheless, the point is to respect
Hypothesis~\ref{HYP-no-simult} rigorously and rule out simultaneity
altogether. Thus, for any couple of events that are considered, one event must
necessarily occur \emph{before} the other. Therefore, either the change of
states of a gene and that of its regulating protein must be considered as one
and only event, or the changes must be separated in time by a delay, possibly
very small. We choose the second solution (by identifying gene state changes and
protein state changes, the first solution demands too severe renouncements of
the original modelling). Thus, we denote by $\delPG[j]{i}$ the time it takes
between a change of states of protein $P_i$ and the according change of states
of a gene $G_j$ which is regulated by it. And we do not exclude the possibility
that the new delays $\delPG[j]{i}$ be evidently negligible in comparison to the
first delays introduced, $\delup$ and $\deldwn$.\smallskip

\noindent Let us concentrate on the \BAN{} of Examples~\ref{EX-contrex}
and~\ref{EX-contrex-newconfig} and let us consider this network as a model of a
genetic system with two genes $G_0$ and $G_1$ and their two protein products
$P_0$ and $P_1$.  Under the three original
hypotheses~\ref{HYP-asynchrone},~\ref{HYP-same-transition-time}
and~\ref{HYP-fortuitintheface}, the behaviour of this network is given by the
\TG{s} of Examples~\ref{EX-contrex} and~\ref{EX-contrex-newconfig}. Under the
additional but necessary Hypothesis~\ref{HYP-no-simult}, no network
configurations (as defined by Equation~\ref{EQ-newconfig}) are excluded any
longer. This implies in particular that Hypothesis~\ref{HYP-no-simult} (and
\emph{a fortiori} Hypothesis~~\ref{HYP-asynchrone}) and
Hypothesis~\ref{HYP-fortuitintheface} cannot strictly be satisfied at once. And
since transitions still are asynchronous, in theory, in our example network of
size two, four elementary transitions must be considered in every configuration
(some of which may be null):\\[2mm] \centerline{
  \begin{tikzpicture}[descr/.style={fill=white,inner sep=2.5pt}] 
    \node at (0,0) (xg) {$\PG{x}{g}=\PGPG{x_0}{g_0}{x_1}{g_1}$}; 
    \node at (-6,0.4) (GH) {$\PGPG{g_0}{g_0}{x_1}{g_1}$}; 
    \node at (-3.5,-0.5) (GB) {$\PGPG{x_0}{f_0(x)}{x_1}{g_1}$}; 
    \node at (3.5,-0.5) (DB) {$\PGPG{x_0}{g_0}{x_1}{f_1(x)}$}; 
    \node at (6,0.4) (DH) {$\PGPG{x_0}{g_0}{g_1}{g_1}$}; 
    \path[-open triangle 60] (xg) edge (GH) edge (GB) edge (DH) edge (DB);
  \end{tikzpicture}
} 
Rather than giving the complete new version of the \TG{} of this network
(which contains $16$ configurations), let us suppose that $\delup[0] <
\delup[1]$ and let us consider the diagram pictured in
Figure~\ref{FIG-contrex}. It describes informally four alternate scenarios that
can occur under the current general Hypotheses~\ref{HYP-asynchrone},~
\ref{HYP-same-transition-time} and~\ref{HYP-no-simult}. All four of these
scenarios start with the activation of $P_1$ in the network configuration
modelling the situation in which both genes are active but unexpressed. The
system is supposed to be isolated and free of any exterior perturbations so that
failures in protein concentrations (resp. in the expression of genes) due to
other factors than the deactivation of their coding genes (resp. the inhibition
of their regulating proteins) are ignored. In particular, cases in which $g_0 =
0$ are disregarded. Let us comment on this new modelling of the system
behaviour. First, several ``kinds'' of configurations
\begin{equation*}
  \PGPG{1}{1}{1}{1}
\end{equation*}
are involved. There are those in which the signal emitted by protein $P_1$
addressed to its coding gene $G_1$ has not (yet) been received and there are
those in which it has. Figure~\ref{FIG-contrex} suggests that in the latter
configurations, the activation of $G_1$ is more stable than in the former in
which the activation of $G_1$ is re-enforced by the presence of an activating
protein (rather than just being made possible by the absence of an inhibiting
one). Second, besides their respective complexity and the transient trajectories
they involve, Figure~\ref{FIG-contrex} and the \TG{} of
Example~\ref{EX-contrex-newconfig} yield similar results if one focuses on limit
behaviours. These remarks may encourage one in arguing that
Figure~\ref{FIG-contrex} represents a futile complexification of the original
modelling. However, let us recall that, importantly, this new modelling follows
from a rigorous adaptation of the original theory of \BAN{s} with delays and
that it was made necessary to make the theory fit coherently to its priorly
intended interpretation. Moreover, in this context where
Hypotheses~\ref{HYP-asynchrone},~\ref{HYP-same-transition-time}
and~\ref{HYP-no-simult} all need to be respected, the possible duality in the
meaning of network configurations that are formally considered as one unique
configuration, as well as the existence of different intermediary steps leading
to the limit stable configurations are important. Both these features of the
modelling cannot be disregarded \emph{a posteriori} since they result directly
and essentially from the work of theorisation and interpretation that was made
\emph{a priori}. In particular, if the network was actually a sub-network of a
larger network, then some other automata may be affected by the states of the
two automata that are considered. As a consequence, the behaviour of the whole
network could be considerably affected by slight deviations from what is
predicted by the original \TG{}. These deviations could, for instance, consist
in gene $G_1$ being less stable in some configurations where $x_0 = x_1 = g_0 =
g_1$ than in others. Or they could be the intermediary steps through which the
network may pass even if only to remain a fleeting instant. Since our new
formalism is a rigorous refinement and adaptation of the original, the
pertinence of the modelling it produces cannot be questioned unless that of the
original version is. It can, however, be criticised for its increased complexity
all the more so that Figure~\ref{FIG-contrex} proposes an incomplete description
of the network behaviour. Indeed, in Scenario 3, not all signals emitted are
received.  Protein $P_1$ is active during a certain period, \ie, its
concentration in the cell exceeds the threshold for some time and yet, its
coding gene $G_1$ ignores it. For the sake of formal rigour and consistency, the
description should be completed with a sequel to this branch which would
increase further the complexity of the description. Further, to interprete this
new formal description and to gauge the feasibility and plausibility of any of
the scenarios it mentions, relationships of precedence need to be established
between the events that are considered. Consequently, delays (and perhaps
protein concentrations) need to be compared precisely. But this leads to the
conclusion that to exploit the new formalism, it must be refined further still,
suggesting in the end that a continuous framework would be better suited for the
modelling problems at hand here.  Thus, an eventual change of modelling paradigm
seems to be required.\smallskip

\noindent \BAN{s} with delays thus provide indeed a modelling of some real
phenomena that takes time flow into account. The supplementary parameters that
are integrated in these models apparently provide an enhancement of the original
model that potentially allows to better understand the conditions required for
one scenario to occur rather than another according to the delays that separate
the events involved in each.  But the modelling power of this formalism relies
substantially on the non trivial interpretation of some formal hypotheses, in
particular, the absolute non-simultaneity of any (lasting) couple of events and
the assumption that whatever the current state of the cell, a protein always
takes the same amount of time to change states.

\section{Effective modelling and observing of a system}
\label{SEC-effective-modelling}

Once a formal well-bounded framework has been accepted and coherently associated
to a correspondence with reality, the observing and modelling of the behaviour
of a particular system can be carried out. The present section is set in this
general context. More specifically, it lies on the hypothesis that neither the
machinery of a system nor its movements can ever be effectively
observed. Instead, it is the successive punctual positions or states of the
system that can be. As an illustration, on the \iieme{9} of august 2011 at local
time $12:30:56$ in the place with geographic coordinates $(45.662587,1.789201)$,
we may observe that proteins $P_1$ and $P_2$ are both present in very high
concentrations in the liver cells of the rat Bobby. Two minutes later, we may
note that the concentration of $P_1$ is almost null. Further, we may perform the
same experience at different dates, times, places and with different rats, and
make similar observations whereas a second experience in which $P_2$ is
initially absent always yields different observations: the concentrations of
$P_1$ that are recorded always are approximately identical during several
hours. In both experiences, no dynamical phenomenons such as concentration
changes are observed, only their results are. In addition, the influence of
protein $P_2$ on the concentration of protein $P_1$ is not observed effectively
although several situations are observed from which this influence (or perhaps a
different conclusion) may be inferred.  Now, primarily, the very nature of
modelling makes it impossible for any model to even come close to reproducing or
describing a real system faithfully: from the start, the theorisation process
can only aim at designing a framework that hopefully will allow models to
reproduce, describe or explain partially some properties of the system. Here, we
add the hypothesis that the only available information concerning a real system
is derived from necessarily partial observations of its behaviour. In summary,
in our context, the effective modelling of a real system is supposed to start
with a series of observations of its behaviour.  Once a ``reasonable''
collection of observations has been established, it becomes possible to build a
model of the system behaviour. To derive from it a model of the system structure
and underlying mechanisms, however, is a tricky task that necessarily encounters
sources of incompletion which call for supplementary approximations and
hypotheses. And these new approximations and hypotheses have notable
significance with regards to the reality/theory correspondence.

\subsection{From observations of a system to a model of its underlying
  interactions}
\label{SEC-observing}

The very first step in the process of effective modelling consists in
determining a set of formal configurations to model the states in which the
system is observed. Here, we focus on systems that contain a finite number of
interacting elements which are intended to be modelled by \BAN{s} so the formal
configurations that are chosen are Boolean vectors $x \in \Bn$ where each
Boolean coefficient $x_i \in \B$ of $x$ is supposed to represent the state of a
system element identified by the label $i < n$.  The choice of the set of
configurations is crucial because it requires a knowledge of what is the
``interior'' of the system that is modelled and what is its ``exterior''.
Indeed, to specify the size of the model network and decide what is the
dimension $n$ of the Boolean vectors, the number of elements interacting in the
system needs to be known. This often means that the set of elements itself needs
to be known. Therefore, in most cases, from the very start, modelling requires
to assume or to know that the content and the frontiers of the system to be
modelled are exactly those that are suggested by or derived from observations of
it. In other terms, unless certainty may be established on this subject, the
following hypothesis is needed:
\begin{hyp}
  The number $n \in \N$ of interacting automata in the network is known.
  \label{HYP-nomoreautomata}
\end{hyp}
Then, the \emph{movements} or changes of the system can be considered for
modelling. Let us suppose that $x \in \Bn$ is a configuration that models the
system state which has been observed right before the system state modelled by
configuration $y \in \Bn$. Then, the change underwent by the system between
these two observed states is modelled by the couple $(x, y) \in \Bn \times
\Bn$. Such couples are called \emph{observed transitions} and denoted as
follows:
\begin{equation*}
  x\transOBS y\text{.}
\end{equation*}
It is important to note that observed transitions are not necessarily network
transitions. This will become clear in the sequel. The set of all observed
transitions derived from observations of a system defines an \emph{observed
  \TG{}} whose role is precisely to model the system behaviour.\smallskip

\noindent With an observed \TG{} \T{obs} in hand, the modelling can be pushed a
step further. The aim is now to infer from \T{obs} knowledge concerning the
underlying mechanisms that contribute in producing the events that are described
in \T{obs}. In short, this third step of modelling aims at determining the
causes of the effects observed. Thus, the information carried by \T{obs} must be
analysed and exploited in order to define a set of \LTF{s} $\F$ and a network
structure $\G$ which must both be coherent with \T{obs}, with the hypotheses
that were made beforehand, with the underlying theory, and with its
correspondence with reality, pre-defined by the prior work of theorisation.  A
natural way to do this is to derive $\F$ using one of
Equations~\ref{EQ-GTtofiPAR},~\ref{EQ-GTtofiSEQ} and~\ref{EQ-GTtofi} or to
exploit Algorithm~\ref{ALGO-BS} and then derive $\G$ from $\F$. However, as
mentioned in Section~\ref{SEC-infer}, this method requires some additional
information and failing which, some additional hypothesis concerning the
\emph{nature} of the transitions in \T{obs}. As a consequence, if $\F$ and $\G$
are defined this way and if they are effectively intended to be used, studied
or exploited to derive additional information on the original system, then the
following question needs to be answered: how can $\F$ and $\G$ be interpreted,
\ie, what modelling power can they be granted? In the absence of a complete and
satisfiable answer to this question some new hypotheses need to be
made.\smallskip

\begin{figure}[h!]  
  \centerline{ 
    \begin{tabular}{c@{\hspace{3cm}}c} 
      \scalebox{0.71}{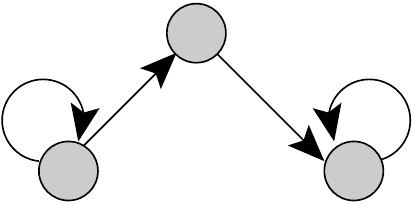}& 
      \scalebox{0.71}{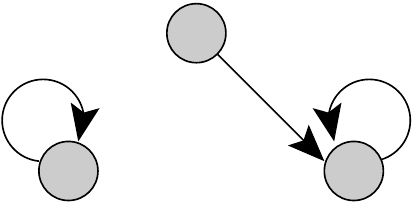} \\
      $a.$ & $b.$
    \end{tabular}  
  }
  \caption{Possible interaction graphs of Boolean automata network whose
    transition graph \T{obs} presented in Example~\ref{EX-HYP-nomoreautomata}.}
  \label{FIG-struct-original-HYP-nomoreautomata}
\end{figure}

\noindent The rest of this section proposes to illustrate some of these
hypotheses and discuss their meaning and impact. In the examples that are taken,
for the sake of their simplicity, Boolean automata networks serve as models of
Boolean automata networks.  In other terms, the networks play both the parts of
observed system to be modelled and model of the observed system. One might
imagine that a computer program simulates the behaviour of a Boolean automata
network $N$ and prints out on a monitor the states it takes. Of course, when the
program is executed, on the one hand, the program may be designed to hide some
parts of the information and on the other hand, our diligence in watching the
screen may not be perfect. We may, for instance, only look periodically at it,
or from time to time at random, or we may perhaps forget to look at it
altogether. In any case, the lists of successive configurations that are
observed yield a transition graph \T{obs} from which, as mentioned above, sets
of automata, of interactions and of local transition functions may be inferred
to define a new Boolean automata network $N'$ supposed to model the original
network $N$.
\begin{example} 
  \label{EX-HYP-nomoreautomata}
  Let us first illustrate the importance of
  Hypothesis~\ref{HYP-nomoreautomata}. To do so, let us suppose that the
  observed transition graph \T{obs} contains only the two following transitions:
  \begin{equation*}
    (1,0) \transOBS (1,1) \quad\text{and}\quad (0,0) \transOBS (0,1)\text{.}
  \end{equation*}
  By Hypothesis~\ref{HYP-nomoreautomata}, we can start by deriving that the set
  of configurations of the observed network $N$ is $\B^2$ and that it involves
  two automata which are referred to here as automaton~$0$ and
  automaton~$1$. Choosing to model these two automata by two automata with the
  same names and using Equations~\ref{EQ-GTtofi} or~\ref{EQ-GTtofiSEQ} (\ie,
  assuming transitions of \T{obs} are elementary) yields the following \LTF{s}:
  \begin{equation*}
    f_0' : x \in \B^2 \mapsto x_0 \quad\text{and}\quad 
    f_1' : x \in \B^2 \mapsto 1\text{.}
  \end{equation*}
  The resulting model $N'$ of $N$ has the same behaviour as $N$ (its set of
  possible transitions is exactly those of \T{obs}). It has $\F = \{f_0',f_1'\}$
  as set of \LTF{s} and the digraph pictured below as interaction
  structure:\\[2mm]
  \centerline{
    \scalebox{0.71}{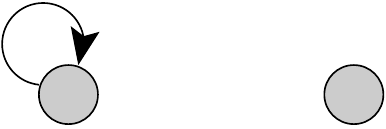}
  }
  If Hypothesis~\ref{HYP-nomoreautomata} were not satisfied, however, $N$ could
  contain a third automaton, automaton~$2$, that observations of $N$ give no
  evidence of. Then, when transition $(1,0) \transOBS (1,1)$ is observed, it
  could be that what is really occurring is a series of transitions that has the
  same effect of increasing the state of automaton~$1$ but involve the hidden
  automaton~$2$:
  \begin{equation*}
    (1,0,0) \transRT (1,0,1) \transRT (1,1,1)\text{,}
  \end{equation*} 
  or perhaps just:
  \begin{equation*}
    (1,0,1) \transRT (1,1,1)\text{.} 
  \end{equation*}
  In that case, according to whether automaton~$0$ activates automaton~$2$ or
  not, or according to whether $(1,0,0) \transRT (1,0,1)$ is possible or not,
  the \LTF{} of automaton~$2$ could equal:
  \begin{equation*}
    f_2: x \in \B^3 \mapsto x_0 \quad\text{or}\quad 
    f_2: x \in \B^3 \mapsto 0\text{.}
  \end{equation*}
  That of automata~$0$ and~$1$ could respectively equal: 
  \begin{equation*}
    f_0: x \in \B^3 \mapsto {x_0} \quad\text{and}\quad
    f_1: x \in \B^3 \mapsto {x_1\vee x_2}\text{.}
  \end{equation*}
  As a consequence, rather than the interaction graph of $N'$ that suggests that
  automata~$0$ and~$1$ are independent and depend only on themselves, network
  $N$ could instead have one of the interaction graphs pictured in
  Figure~\ref{FIG-struct-original-HYP-nomoreautomata}.\eex
\end{example}

\noindent Example~\ref{EX-HYP-nomoreautomata} shows that relying on
Hypothesis~\ref{HYP-nomoreautomata} in the construction of a model $N'$ of a
system $N$ may possibly be responsible for the inferring of a structure that
differs significantly from the reality of $N$. This way,
Hypothesis~\ref{HYP-nomoreautomata} may cause to mistaken the cause of an event
that is observed (the change of state of automaton $1$ in
Example~\ref{EX-HYP-nomoreautomata}, for instance). Now, let us consider some
new hypotheses that apply more specifically to the transitions of
\T{obs}.\medskip
\begin{figure}[t!]
  \centerline{
    \begin{tabular}{cc}
      \begin{minipage}{4cm}
        \vspace{-4cm}\scalebox{0.8}{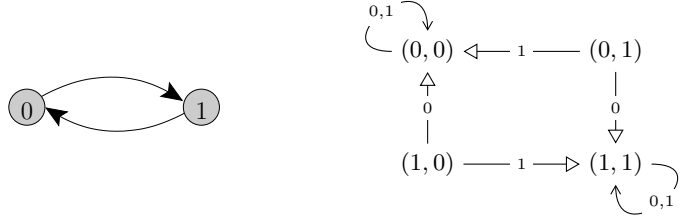}
      \end{minipage} &
      \begin{tikzpicture}[descr/.style={fill=white,inner sep=2.5pt}]
        \node at (0,1.5)   (zero)  {$(0,0)$}; 
        \node at (2.5,1.5) (un)    {$(0,1)$};           
        \node at (0,0)   (deux)  {$(1,0)$}; 
        \node at (2.5,0) (trois) {$(1,1)$};   
        \path (un) edge[-open triangle 60] node[descr] {$\scriptstyle 1$} 
        (zero) edge[-open triangle 60] node[descr] {$\scriptstyle 0$} (trois);
        \path (deux) edge[-open triangle 60] node[descr] {$\scriptstyle 0$} 
        (zero) edge[-open triangle 60] node[descr] {$\scriptstyle 1$} (trois);
        \draw (trois) to [->, >=angle 60, min distance=10mm, in =-90, out=0] 
        node[descr] {$\scriptstyle 0,1$} (trois);
        \draw (zero) to [->, >=angle 60, min distance=10mm, in =90, out=180] 
        node[descr] {$\scriptstyle 0,1$} (zero);
      \end{tikzpicture}
    \end{tabular}}\vspace*{-0.6cm} 
  \caption{Interaction graph and \SSIG{} of the Boolean automata network of
    Example~\ref{EX-HYP-noexterior}.}
  \label{FIG-HYP-noexterior} 
\end{figure}

\noindent First, Hypothesis~\ref{HYP-noexterior} below allows to focus
effectively on \T{obs} and on any information (such as $\G$ and $\F$) that can
be drawn solely from it as suggested above. Indeed, it assumes that no exterior
force or reason is responsible even partially for an observation that is made:
\begin{hyp} 
  Every behaviour of the system is caused by factors that are contained in its
  own definition.
  \label{HYP-noexterior}
\end{hyp}
\begin{example} 
  Let us suppose that the observed \BAN{} $N$ satisfies the following
  properties. It contains two automata, automaton $0$ and automaton $1$ whose
  \LTF{s} are defined by:
  \begin{equation*}
    \forall x\in\B^2,\ f_0(x) = x_1 \quad\text{and}\quad f_1(x) = x_0\text{.}
  \end{equation*}
  Its interaction graph is the digraph pictured on the left of
  Figure~\ref{FIG-HYP-noexterior} and, for some reason, $N$ can only perform
  asynchronous transitions and it can potentially perform them all. Its
  behaviour, in the absence of any exterior perturbations is described by the
  \SSIG{} on the right of Figure~\ref{FIG-HYP-noexterior}.\smallskip

  \noindent Now, let us suppose that $N$ is subjected to some exterior forces
  that impose restrictions on its behaviour and that these restrictions can be
  translated in terms of updating constraints. For instance, it could be that
  active automata tend to be updated before inactive automata. In that case,
  transitions $(0,1) \Seq[0] (1,1)$ and $(1,0) \Seq[1] (1,1)$ in the \SSIG{} of
  Figure~\ref{FIG-HYP-noexterior} may be considered as unlikely or
  impossible. Here, let us suppose differently that an exterior force causes
  automaton $0$ to be much faster in switching states than automaton $1$
  is. Thus, $N$ behaves as if it were obeying to the sequential \us{} $\delta
  \equiv \{0\}\{1\}$. An outside observer that is aware that $N$ only performs
  asynchronous transitions might possibly ignore that $N$ is submitted to any
  updating constraints and as a consequence might record the following \TG{}
  \T{obs}:\\[2mm] \centerline{
    \begin{tikzpicture}[descr/.style={fill=white,inner sep=2.5pt}]
      \node at (0,1.5)   (zero)  {$(0,0)$};
      \node at (2.5,1.5) (un)    {$(0,1)$};
      \node at (0,0)   (deux)  {$(1,0)$};
      \node at (2.5,0) (trois) {$(1,1)$};
      \path (un) edge[-open triangle 60] node[descr] {$\scriptstyle 0$} (trois);
      \path (deux) edge[-open triangle 60] node[descr] {$\scriptstyle 0$} 
      (zero);
      \draw (trois) to [->, >=angle 60, min distance=10mm, in =-90, out=0] 
      node[descr] {$\scriptstyle 0,1$} (trois);
      \draw (zero) to [->, >=angle 60, min distance=10mm, in =90, out=180] 
      node[descr] {$\scriptstyle 0,1$} (zero);
    \end{tikzpicture}
  }
  \noindent If this were the case, the model $N'$ of $N$ would have \LTF{s}
  defined by:
  \begin{equation*}
    \forall x \in \B^2,\ f_0'(x) = f_1'(x) = x_1
  \end{equation*} 
  and the following structure:\\[2mm]
  \centerline{ 
    \scalebox{0.8}{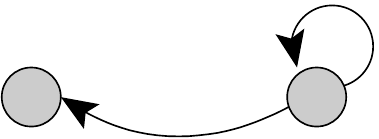}
  } 
  The influence that automaton $0$ has on automaton $1$ would thus not be
  revealed at all because of the precedence of $0$-updates over $1$-updates.\eex
  \label{EX-HYP-noexterior}
\end{example}

\noindent The example above highlights the non-triviality of choosing to ignore
the possibility that the network behaviour might be subjected to exterior
influences that cannot be formalised as proper automata influences.  In this
example, focus on the particular but important case of influences that can be
taken into account through the defining of an \us{}.\medskip

\noindent The next hypothesis specifies that every system behaviour that should
have been observed, with respect to the ``observation protocol'', has indeed
been observed and modelled in \T{obs}.
\begin{hyp} With respect to the nature of observed transitions, \T{obs} contains
  all paths and behaviours of the system model.
  \label{HYP-alltrans}
\end{hyp}
Thus, if the system is supposed to have been observed often enough so as to
witness each of its least changes then, by Hypothesis~\ref{HYP-alltrans},
\T{obs} is the \GIG{} or the \GGIG{} of its model. If, on the contrary, the
system is not supposed to have been observed often enough to acknowledge every
elementary transition, then, Hypothesis~\ref{HYP-alltrans} imposes that \T{obs}
still contain a ``representation'' of any possible system behaviour. If
transition $x \transRT y$ is a (possibly elementary) transition that models a
system change of states, then, by Hypothesis~\ref{HYP-alltrans}, it must be
taken into account: the path that involves this transition must be represented
in \T{obs}, even if not elementarily. In some sense,
Hypothesis~\ref{HYP-alltrans} is a stronger version of
Hypothesis~\ref{HYP-noexterior} that assumes that all behaviours that an
observer has no knowledge of are impossible under all circumstances. Conversely,
it means that any change that is not observed is not possible, \ie, any
transition that does not belong to \T{obs} either is a contained in a path of
\T{obs} or is impossible altogether. As a consequence,
Hypothesis~\ref{HYP-alltrans} implies Hypothesis~\ref{HYP-fixity} below which
relates (if not equates) the exercise that consists in observing movement to
that of observing fixity.
\begin{hyp} 
  A configuration with out-degree null in \T{obs} models a stable configuration
  of the system.
  \label{HYP-fixity}
\end{hyp}
\begin{example}
  Let us suppose that only one transition of the \BAN{} $N$ has been observed:
  \begin{equation*}
    (1,0) \transOBS (1,1)\text{.}
  \end{equation*}
  Using Equation~\ref{EQ-GTtofiSEQ}, we derive a model network $N'$ of size $2$
  with the following \LTF{s}:
  \begin{equation*}
    f_0': \fcn{\B^2}{\B}{x}{x_0} \quad\text{and}\quad 
    f_1': \fcn{\B^2}{\B}{x}{x_0 \vee x_1}\text{.}
  \end{equation*}
  The interaction structure of $N'$ is:\\[2mm]
  \centerline{
    \scalebox{0.71}{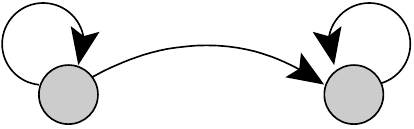}
  }
  Supposing that Hypothesis~\ref{HYP-alltrans} is actually not satisfied
  because, for instance, transition $(0,1) \transRT (1,1)$ has been missed, it
  could be that the local transition function of automaton $0$ in $N$ is in
  reality $f_0 : x \mapsto x_0 \vee x_1$ rather than function $f_0'$. And the
  structure of $N$ rather looks like:\\[2mm]
  \centerline{
    \scalebox{0.71}{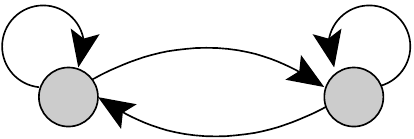}
  }
  than like the structure of the model $N'$. Missing some observations leads
  here to miss some interactions. Conversely, missing some observations can lead
  to infer interactions that do not exist. Indeed, this happens if network $N$
  can perform transition $(0,0) \transRT(0,1)$ although it has not been
  observed. If this is the case, the \LTF{} of automaton $1$ in $N$ does not
  equal $f_1'$ either but, instead, the constant function $f_1 : x \mapsto 1$
  and the structure of $N$ is as follows:\\[2mm]
  \centerline{ \scalebox{0.71}{\input{struct2.pdf_t}} }
  Thus, in reality, automaton $1$ of $N$ could be submitted to no influence from
  automaton $0$ contrary to what is suggested by model $N'$. Further, combining
  this latter example to Example~\ref{EX-HYP-nomoreautomata}, we find that it
  could also be that $N$ has a structure that looks like that of
  Figure~\ref{FIG-struct-original-HYP-nomoreautomata}~$a.$ (this could happen if
  automaton $2$ was not observed and transition $(0,0) \transRT (0,1)$ was
  missed).\eex
  \label{EX-HYP-alltrans}
\end{example}

\noindent The examples above highlight how failing to observe some of the
possible movements of the system may hinder the correct reconstruction of its
set of interactions by either leading to the inference of non-existing
interactions or missing some existing ones.\medskip

\noindent In some cases, to exploit \T{obs}, one may hypothesise on the
granularity of events as in the last two hypotheses we mention here.
\begin{hyp} 
  \T{obs} is an elementary \TG{}.
  \label{HYP-elementary}
\end{hyp}
\begin{example} Let \T{obs} be the following digraph:\\[2mm]
  \centerline{
    \begin{tikzpicture}
      \node at (0.2,0) (zero) {$(0,0)$}; 
      \node at (2.6,0) (trois) {$(1,1)$} ; 
      \node at (6.2,0) (un) {$(0,1)$} ;           
      \node at (8.6,0) (deux) {$(1,0)$} ;     
      \draw[-, bend right=30,](zero) -- (0.9,0.3);
      \draw[->, >=angle 60, bend right=30,](1.8,0.3) -- (trois);
      \draw[-, snake=bumps,segment aspect=-2,](0.9,0.3) --(1.8,0.3);
      \draw[->, >=angle 60, bend right=30,] (0.9,-0.3)-- (zero);
      \draw[-,  bend right=30,](trois) -- (1.8,-0.3);
      \draw[-, snake=bumps,segment aspect=-2,](1.8,-0.3) -- (0.9,-0.3);
      \draw[-, bend right=30,](un) -- (6.9,0.3);
      \draw[->, >=angle 60, bend right=30,](7.8,0.3) -- (deux);
      \draw[-, snake=bumps,segment aspect=-2,](6.9,0.3) --(7.8,0.3);
      \draw[->, >=angle 60, bend right=30,] (6.9,-0.3)-- (un);
      \draw[-,  bend right=30,](deux) -- (7.8,-0.3);
      \draw[-, snake=bumps,segment aspect=-2,](7.8,-0.3) -- (6.9,-0.3);
    \end{tikzpicture}
  } 
  Assuming all transitions in \T{obs} are elementary and using
  Equation~\ref{EQ-GTtofi}, a model $N'$ may be derived with the following
  \LTF{s}: 
  \begin{equation*}
    f_0': x \in \B^2 \mapsto {\neg x_0} \quad\text{and}\quad 
    f_1': x \in \B^2 \mapsto {\neg x_1}\text{.}
  \end{equation*}
  and the structure pictured below:\\[2mm]
  \centerline{
    \scalebox{0.71}{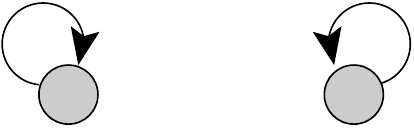}
  }
  According to this model, the two automata of $N$ do not interact. However, let
  us suppose that \T{obs} has been derived by observing the configuration of $N$
  only once per time unit, while, in reality, the network $N$ performs two
  transitions per time unit. Contrary to Hypothesis~\ref{HYP-elementary},
  \T{obs} is not elementary. Further, let us suppose that the synchronous
  transitions that are observed in \T{obs} are actually due to a series of two
  asynchronous transitions so that the \SSIG{} of $N$ is:\\[2mm]
  \centerline{
    \begin{tikzpicture}[descr/.style={fill=white,inner sep=2.5pt}]
      \node at (0,1.5) (zero) {$(0,0)$}; 
      \node at (2.5,1.5) (un) {$(0,1)$} ;           
      \node at (0,0) (deux) {$(1,0)$} ; 
      \node at (2.5,0) (trois) {$(1,1)$} ;   
      \path (zero) edge[-open triangle 60] node[descr] {$\scriptstyle 1$} (un);
      \path (un) edge[-open triangle 60] node[descr] {$\scriptstyle 0$} (trois);
      \path (trois) edge[-open triangle 60] node[descr] {$\scriptstyle 1$} (deux);
      \path (deux) edge[-open triangle 60] node[descr] {$\scriptstyle 0$} (zero);
    \end{tikzpicture}
  } 
  and its structure is the digraph below:\\[2mm]
  \centerline{
    \begin{minipage}{0.02\textwidth}
      ~~~
    \end{minipage}
    \begin{minipage}{0.95\textwidth}
      \centerline{
        \scalebox{0.71}{\input{struct3a.pdf_t}}
      }
    \end{minipage}
    \begin{minipage}{0.02\textwidth}
      \vspace*{2mm}\eex
    \end{minipage}
  } 
  \label{EX-HYP-elementary1}
\end{example}

\noindent Example~\ref{EX-HYP-elementary1} shows that assuming that observed
transitions are elementary is not a negligible choice either, in terms of
modelling. In the same lines, the next example shows that this is true even when
transitions seem to be asynchronous.
\begin{example}
  Let $N$ be of size $3$ and let us suppose that its behaviour has been observed
  starting in each of its $8$ different configurations. As a result of this
  experimentation the following list of possible transitions has been
  obtained:\\[2mm]
  \centerline{
    \begin{tikzpicture}
      \node at (3,1) (zero) {$(0,0,0)$}; 
      \node at (6,0) (un) {$(0,0,1)$} ;           
      \node at (0.5,1) (deux) {$(0,1,0)$} ; 
      \node at (8.5,0) (trois) {$(0,1,1)$} ;   
      \node at (8.5,1) (quatre) {$(1,0,0)$}; 
      \node at (0.5,0) (cinq) {$(1,0,1)$} ;           
      \node at (6,1) (six) {$(1,1,0)$} ; 
      \node at (3,0) (sept) {$(1,1,1)$} ;
      \draw[-, snake=bumps,segment aspect=-2,](deux)--(2.1,1);
      \draw[->, >=angle 60](2.1,1)--(zero);
      \draw[-, snake=bumps,segment aspect=-2,](cinq)--(2.1,0);
      \draw[->, >=angle 60] (2.1,0)--(sept);
      \draw[-, snake=bumps,segment aspect=-2,](six)--(7.6,1);
      \draw[->, >=angle 60](7.6,1)--(quatre);
      \draw[-, snake=bumps,segment aspect=-2,](un)--(7.6,0);
      \draw[->, >=angle 60](7.6,0)--(trois);
    \end{tikzpicture}
  } 
  Assuming, by Hypothesis~\ref{HYP-elementary} that all four of these
  transitions are elementary (and asynchronous), a model $N'$ can be derived
  with the following \LTF{s}:
  \begin{equation*}
    f'_0: x \in \B^3 \mapsto {x_0} \text{,}\quad 
    f'_1: x \in \B^3 \mapsto {x_1 \vee x_2} \quad\text{and}\quad 
    f'_2: x \in \B^3 \mapsto {x_2}\text{.}
  \end{equation*}
  and the following interaction graph:\\[2mm]
  \centerline{
    \scalebox{0.8}{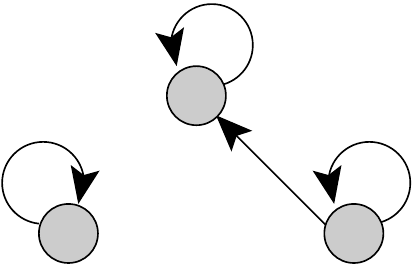}
  }
  This modelling of $N$ by $N'$ suggests that the only interaction existing
  between two different automata of the network consists in the activation of automaton
  $1$ by automaton $2$. Supposing again that Hypothesis~\ref{HYP-elementary} is
  not satisfied, it might be that $N$ is actually updated with the \us{} $\delta
  \equiv \{0,1\}, \{1,2\}, \{0,2\}$ of period $3$. In that case, the four
  apparently asynchronous transitions that are observed are actually rough
  approximations of the four sequences of transitions that figure below. As
  before, this loss of information may be explained for instance by the fact
  that only one observation per time unit was made while $N$ performed $3$
  transitions per time unit.\\[2mm]
  \centerline{
    \begin{tikzpicture}[descr/.style={fill=white,inner sep=2.5pt}]
      \node at (0,0) (cinq) {$(1,0,1)$} ; 
      \node at (0,1.3) (deux) {$(0,1,0)$} ; 
      \node at (2.3,0) (sept) {$(1,1,1)$} ;   
      \node at (2.3,1.3) (zero) {$(0,0,0)$}; 
      \node at (5,0) (un) {$(0,0,1)$} ;   
      \node at (5,1.3) (six) {$(1,1,0)$} ; 
      \node at (7.3,0) (cinqB) {$(1,0,1)$} ; 
      \node at (7.3,1.3) (deuxB) {$(0,1,0)$} ; 
      \node at (9.6,1.3) (unB) {$(0,0,1)$} ;   
      \node at (9.6,0) (sixB) {$(1,1,0)$} ; 
      \node at (11.9,0) (trois) {$(0,1,1)$} ;   
      \node at (11.9,1.3) (quatre) {$(1,0,0)$}; 
      \path (deux) edge[->, >=angle 60] node[above] {$\scriptstyle \{0,1\}$}  
      (zero);
      \path (cinq) edge[->, >=angle 60] node[above] {$\scriptstyle \{0,1\}$}  
      (sept);
      \path (six) edge[->, >=angle 60] node[above] {$\scriptstyle \{0,1\}$}  
      (deuxB);
      \path (deuxB) edge[->, >=angle 60] node[above] {$\scriptstyle \{1,2\}$}  
      (unB);
      \path (unB) edge[->, >=angle 60] node[above] {$\scriptstyle \{0,2\}$}  
      (quatre);
      \path (un) edge[->, >=angle 60] node[above] {$\scriptstyle \{0,1\}$}  
      (cinqB);
      \path (cinqB) edge[->, >=angle 60] node[above] {$\scriptstyle \{1,2\}$}  
      (sixB);
      \path (sixB) edge[->, >=angle 60] node[above] {$\scriptstyle \{0,2\}$}  
      (trois);
      \draw (zero) to [->,>=angle 60,min distance=10mm,in =45, out=-45] 
      (zero);
      \draw (sept) to [->,>=angle 60, min distance=10mm, in =45, out=-45] 
      (sept);
      \draw (quatre) to [->,>=angle 60,min distance=10mm,in =45, out=-45] 
      (quatre);
      \draw (trois) to [->,>=angle 60,min distance=10mm,in =45, out=-45] 
      (trois);
    \end{tikzpicture}
  } 
  It can be checked that in this case, the local transition functions of the
  three automata of $N$, rather than being equal to the functions $f_i'$ of the
  model $N'$, could equal the following
  \begin{equation*}
    \forall x\in\B^3,\quad f_0(x) = x_2,\quad f_1(x) = x_0,\quad f_2(x) = 
    x_1\text{,}
  \end{equation*}
  so that the interaction structure of $N$ would be a circuit:\\[2mm]
  \centerline{
    \begin{minipage}{0.02\textwidth}
      ~~~
    \end{minipage}
    \begin{minipage}{0.95\textwidth}
      \centerline{
        \scalebox{0.8}{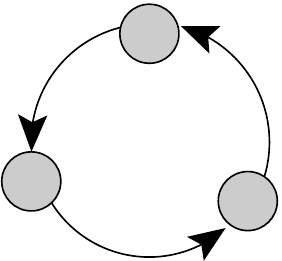}
      }
    \end{minipage}
    \begin{minipage}{0.02\textwidth}
      \vspace*{1.75cm}\eex
    \end{minipage}
  }
  \label{EX-HYP-elementary2}
\end{example}

\noindent Thus, as demonstrated by Example~\ref{EX-HYP-elementary2}, even an
apparently asynchronous transition can in reality be a series of synchronous
transitions. Let us note that generally, when \T{obs} equals a \TG{} of the form
$\T{\delta}$ where $\delta$ is a \BSus{}, then, Equation~\ref{EQ-GTtofiPAR}
yields a model $N'$ in which, informally, all the information of $N$ has been
``condensed'' due to the application of $\delta$ and to the transitivity of
sequences of interactions~\cite{Robert1986,Goles2010,Goles2011}.  As mentioned
in Section~\ref{SEC-infer}, however, the condensed version of the information
cannot be unravelled unless $\delta$ is known precisely (see
Algorithm~\ref{ALGO-BS}).\medskip

\noindent Finally, when, contrary to the previous paragraph, \T{obs} is not
supposed to be elementary (for instance, because a specific \us{} is assumed)
then, the information on the system carried by \T{obs} is incomplete unless the
series of system changes that occur between two successive observations of the
system can be reconstructed.
\begin{hyp} 
  The additional information required to break down each non-elementary
  transition of \T{obs} into an elementary path is known.
  \label{HYP-betweentrans}
\end{hyp}
Hypothesis~\ref{HYP-betweentrans} resembles Hypothesis~\ref{HYP-elementary} but
is stronger. While Hypothesis~\ref{HYP-elementary} assumes that there is nothing
more to each transition than what is observed, Hypothesis~\ref{HYP-betweentrans}
assumes knowledge of what more there is to the transitions that are observed and
considered non-elementary.
\begin{example} 
  Let \T{obs} be the following observed \TG{}:\\[2mm]
  \centerline{
    \begin{tikzpicture}
      \node at (0,0) (zero) {$(0,0)$};       
      \node at (2,-0.5) (deux) {$(1,0)$} ; 
      \node at (4,0) (trois) {$(1,1)$} ;   
      \draw[-, bend right=30,](zero) -- (0.9,0.3);
      \draw[-, snake=bumps,segment aspect=-2,](0.9,0.3) --(3,0.3);
      \draw[->, >=angle 60, bend right=30,](3,0.3) -- (trois);
      \draw[-, snake=bumps,segment aspect=-2,] (zero) -- (1.1,-0.3); 
      \draw[->, >=angle 60,](1.1,-0.3) -- (deux);
      \draw[-, snake=bumps,segment aspect=-2,] (deux) -- (3.2,-0.3); 
      \draw[->, >=angle 60,](3.2,-0.3) -- (trois);
    \end{tikzpicture}
  } 
  where the bottom two transitions are supposed to be elementary and the
  remaining transition $(0,0) \transOBS (1,1)$ is supposed to be an
  approximation of the elementary path $(0,0) \transOBS (1,0) \transOBS
  (1,1)$. Such hypotheses might for instance be made naturally if $N$ is
  considered unable of synchronicity and if transition $(0,1) \transRT (1,1)$ is
  supposed to be impossible because it has not been observed. This, with
  Hypothesis~\ref{HYP-alltrans} and Equation~\ref{EQ-GTtofiSEQ}, yields a model
  $N'$ which has the following \LTF{s}: 
  \begin{equation*}
    f_0': x \in \B^2 \mapsto {x_0 \vee \neg x_1} \quad\text{,}\quad
    f_1': x \in \B^2 \mapsto {x_0 \vee x_1}
  \end{equation*}
  and the following interaction graph:\\[2mm]
  \centerline{
    \scalebox{0.8}{\input{struct2b.pdf_t}}
  } 
  Suppose now that Hypothesis~\ref{HYP-betweentrans} is wrongly assumed and all
  transitions observed are not necessarily asynchronous. $N$ is able to
  perform synchronous transitions and in particular, it can perform transition
  $(0,0) \Pll[\{0,1\}] (1,1)$ in one step. Then, it could be that the \LTF{} of
  automaton $1$ is actually: 
  \begin{equation*}
    f_1: x \mapsto 1
  \end{equation*}
  and its interaction structure is:\\[2mm]
  \centerline{
    \scalebox{0.8}{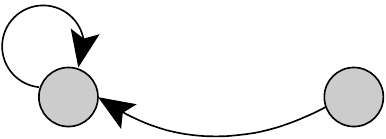}
  }
  Thus, wrongly assuming Hypothesis~\ref{HYP-betweentrans} can also lead to
  deriving the existence of interactions that do not take place in reality.\eex
  \label{EX-HYP-betweentrans}
\end{example}

\noindent To make up for the lack of information, modelling needs to put forward
and rely on some supplementary hypotheses such as
Hypotheses~\ref{HYP-nomoreautomata} to~\ref{HYP-betweentrans}. This proves the
importance of hypotheses. On the one hand, they allow modelling to happen. On
the other hand, as we have endeavoured to show with the examples of this
section, they can mislead it. In particular, they can cause the interaction
structure of a system and of its model to differ significantly, mainly due to the
fact that even in very simple instances of \BAN{s}, there might be different
possible causes to a same effect.

\subsection{Defining networks}
\label{SEC-defining}
\def\defBAN{\defref{DEF-BAN}}

Relying on the previous sections, we now can propose the following formal
definition of \BAN{s}:
\begin{definition}
  A \BAN{} $N$ of size $n \in \N$ is defined by a set of $n$ \LTF{s}:
  \begin{equation*}
    N\ =\ \F = \{f_i\ |\ 0 \leq i<n\}\text{.}
  \end{equation*}
  \label{DEF-BAN}
\end{definition}
Let us note that by Section~\ref{SEC-infer}, from a set $\F$ of \LTF{s}, the
\GIG{} of the network defined by $\F$, can be computed. Conversely, given the
\GIG{} of the network, $\F$ can also be inferred. As a consequence, a \BAN{} is
given equivalently to Definition~\ref{DEF-BAN} by its \GIG{}.

\subsubsection*{Definition~\ref{DEF-BAN} \emph{vs.} the interaction structure}

Let us first recall that with this definition, given a \BAN{} $N = \F$ of size
$n$, the interaction structure of $N$ can be inferred in time ${\cal O}(2^n)$
(see Section~\ref{SEC-infer}). Now, comparing with a possible definition of
\BAN{s} by their interaction structures, we find that, contrary to such a
definition, Definition~\ref{DEF-BAN} implies a first implicit hypothesis:
\begin{hyp} 
  An automaton either obeys \emph{all} of its influences at once (when it is
  updated), or it obeys none.
  \label{HYP-tout-ou-rien}
\end{hyp}
In other words, by our choice of definition, we assume that any updated
automaton $i \in V$ can under no circumstances ignore some of its influences
$(j,i) \in A$.  However, let us note that this restriction can be partially
resolved by a judicious design of the \LTF{s} $f_i$. For instance, if
originally, $f_i(x) = x_j$, then $i$ can be made independent of $j$ when $j$ is
inactive by replacing $f_i$ by $x \mapsto x_j \vee x_i$. This way of making $i$
independent of $j$ can nevertheless only be done relatively to a given network
configuration or set of network configurations\footnote{More generally, let us
  suppose that $f_i(x)$ is given in CNF. For any Boolean formula $\phi(x)$ in
  CNF, let $\phi(x)[0/x_j^\ast]$ equal the formula $\phi(x)$ in which every
  occurrences of both the literal $x_j$ and its negation $\neg x_j$ are replaced
  by $0$ (\eg, if $\phi(x) = \big( x_1 \vee \neg x_2 \vee \neg x_4 \big) \wedge
  \big( x_2 \vee x_3 \big)$, then $\phi(x)[0/x_2^\ast] = \big( x_1 \vee \neg x_4
  \big) \wedge x_3 $). Then, automaton $i$ can be made to ignore the influence
  $(j,i) \in A$ in configuration $y \in \Bn$ if $f_i$ is replaced by the
  function $x \mapsto f_i(x) \vee f_i(y)[0/x_j^{\ast}]$.}.  This leads us to the
following remark. Definition~\ref{DEF-BAN} implies another non-trivial implicit
hypothesis:
\begin{hyp} 
  The result of the interactions that take place in a given configuration
  depends only on that configuration.
  \label{HYP-interactions-fixes}
\end{hyp}
Thus, by Definition~\ref{DEF-BAN}, we assume that given a network interaction
structure $\G = (V,A)$, the nature of the influence represented by the arc
$(j,i) \in A$, when this influence is effective, depends only on the current
network configuration. Consider a network of size $2$, for instance, where
automaton $1$ depends on automaton $0$ and the \LTF{} affected to automaton $1$
is $f_1 : x \in \Bn \mapsto x_0 \in \B$. Then, in configuration $x = (1,0)$
whatever the current environment of the network, if the influence of automaton
$0$ on automaton $1$ does take place (\ie, if automaton $1$ is updated), then
automaton $0$ causes automaton $1$ to take state $1$. Definition~\ref{DEF-BAN}
and more specifically Hypothesis~\ref{HYP-interactions-fixes} disallow any
situation in which the active automaton $0$ deactivates automaton $1$ in
configuration $x$.\smallskip

\noindent In spite of the disputable limitations imposed by
Hypotheses~\ref{HYP-tout-ou-rien} and~\ref{HYP-interactions-fixes}, we still
prefer Definition~\ref{DEF-BAN} to a less restrictive definition of \BAN{s} by
their interaction structures. The reason is mainly arbitrary but motivated,
however, by the belief that the \emph{nature} of each interaction that can
potentially take place in a network (rather than just their existence) is an
essential feature of a network.

\subsubsection*{Definition~\ref{DEF-BAN} \emph{vs.} the \TG{}}

Similarly we have chosen not to define a \BAN{} by a \TG{} (other than its
\GIG{}). In short, the reason for this is to avoid restricting a network to a
system that has only one behaviour. Indeed, if, contrary to
Definition~\ref{DEF-BAN}, a \BAN{} $N$ is defined by a \TG{} $\T{}$ (which is
not the \GIG{} of $N$), then, $N$ can only behave according to $\T{}$. Of
course, this definition may indeed appear better suited for a modelling context
involving steps similar to those described in Section~\ref{SEC-observing} (which
start with the formalisation of an observed system behaviour in terms of a \TG{}
$\T{}$). However, it is important to note again that in such a context, it is
the \emph{transitions} of $\T{}$ rather than the interactions of $N$ that are
initially designed to model a portion of reality. All the examples of
Section~\ref{SEC-observing} show that this difference is significant. The
pertinence of the modelling embodied by these transitions is contingent on the
circumstances in which the real system is observed as well as on \emph{how} it
is observed. In addition, as developed in Section~\ref{SEC-observing}, the
interactions that take place in the network $N$ are only \emph{inferred
  theoretically} based on some non-trivial hypotheses such as
Hypotheses~\ref{HYP-nomoreautomata} to~\ref{HYP-betweentrans}. The meaning, in
terms of modelling, of the interactions derived this way is thus subjected to
the relevance of the hypotheses that are made. However, identifying the observed
\TG{} with the model network $N'$ itself fails at putting forward clearly these
hypotheses which are instead relegated into a pre-accepted definition. As a
consequence, Hypotheses~\ref{HYP-nomoreautomata} to~\ref{HYP-betweentrans}, in
particular, are made implicitly and systematically without being
questioned.\smallskip

\noindent Definition~\ref{DEF-BAN}, on the contrary, either makes irrelevant
Hypotheses~\ref{HYP-nomoreautomata} to~\ref{HYP-betweentrans}, or it requires
that they be made explicit. Indeed, since Definition~\ref{DEF-BAN} is equivalent
to defining a network by its most general behaviour, \ie, its \GIG{}, any
sub-graph of this \TG{} can be considered as a possible elementary behaviour of
the network. Further, the same is true for any other graph whose arcs can be
broken down into paths that belong to the \GIG{}. As a consequence, \BAN{s}
defined by Definition~\ref{DEF-BAN} may be considered as models of systems that
possibly behave differently in different
environments. Hypothesis~\ref{HYP-noexterior}, in particular, is made
superfluous. More generally, given a network $N = \F$ whose interactions are
known \emph{a priori} according to Definition~\ref{DEF-BAN}, in any \TG{}
representing the behaviour of $N$, all transitions have non-ambiguous causes
deriving directly from the network definition
itself. Hypotheses~\ref{HYP-nomoreautomata} to~\ref{HYP-betweentrans} are
irrelevant in this case. And in a converse situation similar to those discussed
in Section~\ref{SEC-observing} in which the interactions of a network are to be
inferred from a partial description of its behaviour, Definition~\ref{DEF-BAN}
makes Hypotheses~\ref{HYP-nomoreautomata} to~\ref{HYP-betweentrans} obviously
needed so they cannot be ignored. The intrinsic ``incompleteness'' of
Definition~\ref{DEF-BAN} (incompleteness in the sense that
Definition~\ref{DEF-BAN} does not specify the network behaviour precisely)
imposes that any supplementary theoretical argument that is used to give a
ruling on the network behaviour be either justified or put forward as purely
hypothetical and convenient.

\section{Discussion}
\label{SEC-discussion}

Modelling generally involves an intermediary step that has not been mentioned in
this paper. This step, purely theoretical, is set between theorisation and
effective modelling. It consists in manipulating objects of the theory alone and
making series of logical inferences to prove or simulate new theoretical
properties that will hopefully be interesting for the rest of the modelling
process. Its main difficulty is to make proper use of the rules of classical
logic commonly used in mathematics. Theorisation and effective modelling, on the
contrary, are ubiquitous steps that require to go back and forth between reality
and theory and that are concerned by any change brought to the theory or to the
set of modelled characteristics of the real system.  Generally, to compare
reality with theory and to elaborate theory on the basis of reality makes
indispensable a thorough and coherent definition and bounding of the modelling
process itself. For this reason, theorisation and effective modelling involve a
considerable difficulty of which the intermediary purely formal step is
immune. This difficulty is to navigate safely but constructively between
experiences and observations of reality on one side, and mathematical
abstractions of it on the other.\smallskip

\noindent Sections~\ref{SEC-theorisation} and~\ref{SEC-effective-modelling} have
both emphasised how the safeguard of consistency throughout the modelling
process is indeed an issue. In the respective contexts of theorisation and
effective modelling, the intrinsic and dual incompleteness of any knowledge as
well as of any representation of reality needs to be overcome. As we have
endeavoured to highlight in this paper, hypothesising is a required but tricky
solution to this. Consider Hypothesis~\ref{HYP-asynchrone}, for
instance. Section~\ref{SEC-delays} argues that no pre-existing interpretation of
the theory of \BAN{s} can serve or yield a valid interpretation for this
hypothesis. Generally, an explanation of a formal choice cannot be drawn solely
from the theory itself (the theory cannot justify itself). But it cannot either
rely exclusively on knowledge or observations of the real system or category of
real systems being modelled. Indeed, arguments drawn from such knowledge and
observations need to be consistent with the pre-defined reality/theory
correspondence. Thus, to justify why, under Hypothesis~\ref{HYP-asynchrone},
certain possible elementary transitions are ignored and more precisely why the
network is supposed \emph{not} to follow the ``maximum speed gradient'', a new
feature of reality must be put forward as argued in
Section~\ref{SEC-delays}. First, a strong supplementary notion of time must be
taken into account. Second, in addition to this, since the fortuitousness of two
events finishing simultaneously is not enough (see Section~\ref{SEC-delays}) to
justify Hypothesis~\ref{HYP-asynchrone} non-ambiguously, one must also consider
an \emph{absolute} notion of non-simultaneity that forbids any two events to
overlap in time. The significance of this can be seen in particular by noticing
that the modelling meaning of the local interaction graphs $\G(x),\, x\in \Bn,$
is quite limited under these assumptions that are intended to support
Hypothesis~\ref{HYP-asynchrone}. Indeed, except for their existing at the same
instant, the interactions in these digraphs are otherwise completely independent.
And this absolute non-simultaneity must then be implanted rigorously into the
pre-defined theory and be connected consistently with the pre-existing
reality/theory correspondence. The refinement represented by
Hypothesis~\ref{HYP-asynchrone} thus obviously calls for another refinement
which is the time notion brought by delays. But, as the end of
Section~\ref{SEC-delays} shows, the consistency requirement also calls for
further refinements. And, in turn, so does the need for non-ambiguous and
plausible predictions. And each refinement adds complexity to the modelling
which increases the difficulty in the interpretation of the theory. This looming
necessity for series of levellings of the approximations that are made by
theorisation runs the risk of ultimately aiming towards a different modelling
paradigm (in the case analysed in Section~\ref{SEC-delays}, the series seems to
aim towards a differential formalism) which would necessarily disavow the
present modelling paradigm by disavowing some of its founding hypotheses (a
continuous formalism obviously denies the founding hypotheses of a discrete
modelling such as a Boolean modelling). Thus, model refining, \ie, the
augmenting of the modelling and interpretation maps associated to a theory (see
Page~\pageref{RTmaps}) as well as augmenting the theory itself is limited on one
hand by the necessity to maintain consistency, and on the other hand by the risk
of increasing the complexity of the modelling and its interpretation to an
extent where a different model is required. The existence of these limits
favours the two following ideas.\smallskip

\noindent First, the ins and outs of a model (the theory and the reality/theory
correspondence) must be understood and fathomed so that no unnecessary
refinement be added. Thus, in a context where duration is not intended to be
modelled specifically, one could exploit the original definition of transitions
of \BAN{s} to model a ``duration-free'' version of delays, bypassing the thorny
problem of simultaneity. Indeed, all elementary transitions that are possible in
an arbitrary configuration $x \in \Bn$ (\ie, all arcs outgoing $x$ in the
\GIG{}) may simply be seen as representing the possibility that in $x$ one or
several automata may end up ulteriorly having changed states. If $i, j \in
\U(x)$ are two automata that can both change states in $x$, then the synchronous
elementary transition $x \Pll[\{i,j\}] \overline{x}^{\{i,j\}}$ does not
necessarily mean that the change of states of $i$ and the change of states of
$j$ have \emph{finished} simultaneously. Actually, it cannot mean this if time
is not modelled (the notion of ``simultaneity'', in particular, is obviously
based on that of time). With the most general and basic interpretation of the
theory of \BAN{s}, synchronous transitions can only be interpreted, ``at the
best'', as models of changes that have been produced during overlapping periods
of time: a transition $x \Pll[\{i,j\}] \overline{x}^{\{i,j\}}$ can be taken to
mean that neither $ \overline{x}^{i}$ nor $ \overline{x}^{j}$ are visited long
enough by the network (if visited at all) for there to occur, in these
configurations, a decisive event that could divert the network from its path
from $x$ to $\overline{x}^{\{i,j\}}$.\smallskip

\noindent The second idea that derives from the evidence of the limits of
refining a model concerns the intrinsic incompleteness of modelling and its
benefit. Levelling the incompleteness of a model to make it apparently better
fit reality can make it hard to bound properly the properties that are supposed
to be modelled and embrace the approximations and hypotheses that were made in
order to do so. Yet, modelling draws its pertinence precisely from its
incompleteness. It is based on approximations of reality that can be very rough,
perhaps even incorrect, and on the formal hypotheses to which they translate
(such as the discreteness of events and time).  Whether these hypotheses be
supported by reasonable arguments or not does not necessarily impact directly on
the pertinence of a model. Indeed, one of the main purposes of hypotheses is to
be questioned and eventually discarded. Convincing model predictions do not
produce an explanation of reality in themselves. An analysis of the hypotheses
on which is based the model, however, may. On one hand, convincing model
predictions suggest that the hypotheses that were chosen are plausible so they
can be used in the elaboration of an explanation (even if only an unrefined
explanation) of the reality observed. On the other hand, the most immediate
cause of a poor simulation of reality by a model can often be found in its
underlying hypotheses which thus gain in being known. In addition, in more
abstract lines, hypotheses are also important because they allow simplifications
that may help formal developments. Furthermore, their very presence carries
questions that can help orient the development and build a mathematical
understanding of the theory that is considered.  Thus, provided they be clearly
expressed, hypotheses are precisely the model features that can be questioned
and consequently that can allow a constructive comparison between reality and
abstraction. In that, they represent one of the essential aspects of modelling,
even (and perhaps especially) when they lack convincing modelling
meaning.\smallskip

\noindent Thus, hypotheses are, we believe, the principal primary foundations of
modelling.  They embody formally the sources of inaccuracy, incompleteness and
possibly incorrectness on which the model is developed and on which relies the
pertinence of its predictions.  A set of ill-identified approximations
represents a risk that the series of interpretations that are drawn from a model
may slightly drift in a way that we have no knowledge of. It may thus be
responsible for an orientation of our understanding of reality that eludes our
control. In addition, more simply perhaps, by nature, hypotheses carry implicit
assertions of facts that are not questioned unless the hypotheses themselves
are. Because we have no guarantee that a question is not pertinent until it
arises and is either answered or discarded deliberately, to avoid implicit
answers, we believe that however subtle or reasonable they may be, hypotheses
should be clearly explicited as well as rigorously analysed, so far as possible.
And in particular, as argued in Section~\ref{SEC-defining}, special attention
should be brought to the design of the theory and to the acknowledgement of the
hypotheses that follow implicitly from its definition choices.\smallskip

\noindent In summary, hypotheses not only are unavoidable embodiments of the
impossibility of modelling a system (rather than a well-bounded set of its
properties), they also are essential. They allow the approximations and
simplifications of reality without which no modelling would be possible. And
since they enclose their intrinsic incompleteness, they allow models to be
effectively informative.

\vspace{1cm}

\subsubsection*{Acknowledgments} We warmly thank K. Perrot for his pertinent
comments. We also thank the \emph{institut rh{\^o}nalpin des syst{\`e}mes
  complexes} (Ixxi) that supported the projet Maajes and the
\emph{r{\'e}seau national des syst{\`e}mes complexes} ({\sc rnsc}) that
supported the project Météding. The present article was partially fostered
by both these projects.


\end{document}